\documentclass[aps,prb,twocolumn]{revtex4-1}
\usepackage[utf8x]{inputenc}
\usepackage{textcomp}
\usepackage{amsmath,amssymb,latexsym,graphicx}
\usepackage{graphics}       
\DeclareGraphicsExtensions{.png,.pdf}
\usepackage{mathtools}
\usepackage{overpic}
\usepackage{color}
\usepackage{subfigure}
\usepackage{natbib}
\usepackage[colorlinks=true,a4paper=true, pdfstartview=FitV,linkcolor=blue, citecolor=blue, urlcolor=blue]{hyperref}
\usepackage[all]{hypcap}
\usepackage{times}
\usepackage{multirow}
\usepackage{bm}
\usepackage{mathrsfs}
\usepackage{relsize}
\newcommand*{\diff}{\mathop{}\!\mathrm{d}}

\def\greendiamond{$\color{green}\diamondsuit$}
\begin{document}
\title{Vortex Structures in Model $p$-Wave Superconducting Sr$_2$RuO$_4$  -- Single 2-Dimensional Band v.s.  Quasi-1-Dimensional Band}
\author{Jia-Wei Huo$^1$, Fu-Chun Zhang$^{1,2}$ }
\affiliation{$^1$Department of Physics, The University of Hong Kong, Hong Kong, China\\
$^2$Department of Physics, Zhejiang University, Hangzhou 310027, China}
\date{\today}
\begin{abstract}
  There have been an interesting debate on the primary source of chiral p-wave superconductivity in Sr$_2$RuO$_4$. We present a comparative study on the vortex structure between a single 2-dimensional (2D) band and quasi-1D band model by using Bogoliubov-de Gennes theory. The pattern of the iso-values  of the local density of state around a vortex has a diamond shape in the quasi-1D model and is much more isotropic in the 2D model.  The spin lattice relaxation rate  well below the superconducting transition temperature is greatly enhanced in the vortex state in the 2D model but not in the quasi-1D model.  These features can be tested by using scanning tunneling microscope and NMR to distinguish the models for the superconductivity in Sr$_2$RuO$_4$.
\end{abstract}
\maketitle
\section{Introduction}
The layered perovskite material Sr$_2$RuO$_4$ has attracted a lot of interests due to the experimental evidence for its spin-triplet superconductivity with broken time-reversal symmetry \cite{Mackenzie2003}. Shortly after the discovery of its superconductivity \cite{Maeno1994}, Rice and Sigrist\cite{Rice1995} and Baskaran\cite{Baskaran1996} pointed out that the superconducting state might be an electronic analogue of the $^3$He-$A$ phase. Within this scenario, assuming the simplest nearest-neighbor pairing interaction, the gap function in terms of the $d$-vector formalism can be expressed compactly as \cite{Miyake1999,Deguchi2004a}
\begin{equation}
  \vec{d}(\bm{k})=\Delta_0\hat{z}(\sin k_x+i\sin k_y).
\end{equation}
Sr$_2$RuO$_4$ is a quasi-2D systems. Its normal state can be well described by multi-orbital band structure with a 2D $\gamma$ band derived from the Ru $d_{xy}$ orbital and two weakly hybridized quasi-1D $\alpha$ and $\beta$ bands derived from Ru $d_{xz}$ and $d_{yz}$ orbitals, whose Fermi surfaces are illustrated in Fig. 2(a). It is natural and has been generally assumed that superconductivity arises primarily on the 2D $\gamma$ band~\cite{Rice1995}, 2D model hereafter. This scenario is consistent with the directional variation of the low temperature specific heat in a magnetic field~\cite{Deguchi2004}.

However, basic questions concerning the primary source of the pairing remain controversial due to important discrepancies between theory and experiment, in spite of intense research work in the past almost two decades~\cite{Kallin2012,Maeno2012}. For example, the 2D model predicts an edge current in the superconducting state, which has not been demonstrated~\cite{Bjornsson2005,Kirtley2007}, although this also has been controversial~\cite{Maeno2012,Raghu2010a,Imai2012}. Very recently, Raghu $et$ $al$. suggested an alternative interesting possibility for the spin-triplet superconductivity in Sr$_2$RuO$_4$~\cite{Raghu2010a,Raghu2012}, where the dominant superconducting instability in the triplet channel occurs on the quasi-1D $\alpha$ and $\beta$ bands, hybridized from the Ru $d_{xz}$ and $d_{yz}$ orbitals. This model (quasi-1D model hereafter) predicts the absence of spontaneous supercurrents at sample edges, contrary to the single 2D band model. One expects an intrinsic anomalous Hall effect due to the multi-band nature of this model~\cite{Taylor2012}, which may explain the observation of nonzero Kerr effect~\cite{Xia2006}. On the other hand, the quasi-1D model would predict a suppression of spin density wave fluctuation at the superconducting transition point~\cite{Huo2013}, which has not been seen in neutron experiments by far~\cite{Braden2002}. Therefore, it is important to explore the possibilities for more experimental consequences within the 2D band or quasi-1D band models and to test the model against experiments.

\begin{figure}[htb!]\centering
  \begin{tabular}{c}
    \resizebox{0.9\linewidth}{!}{
      \begin{overpic}{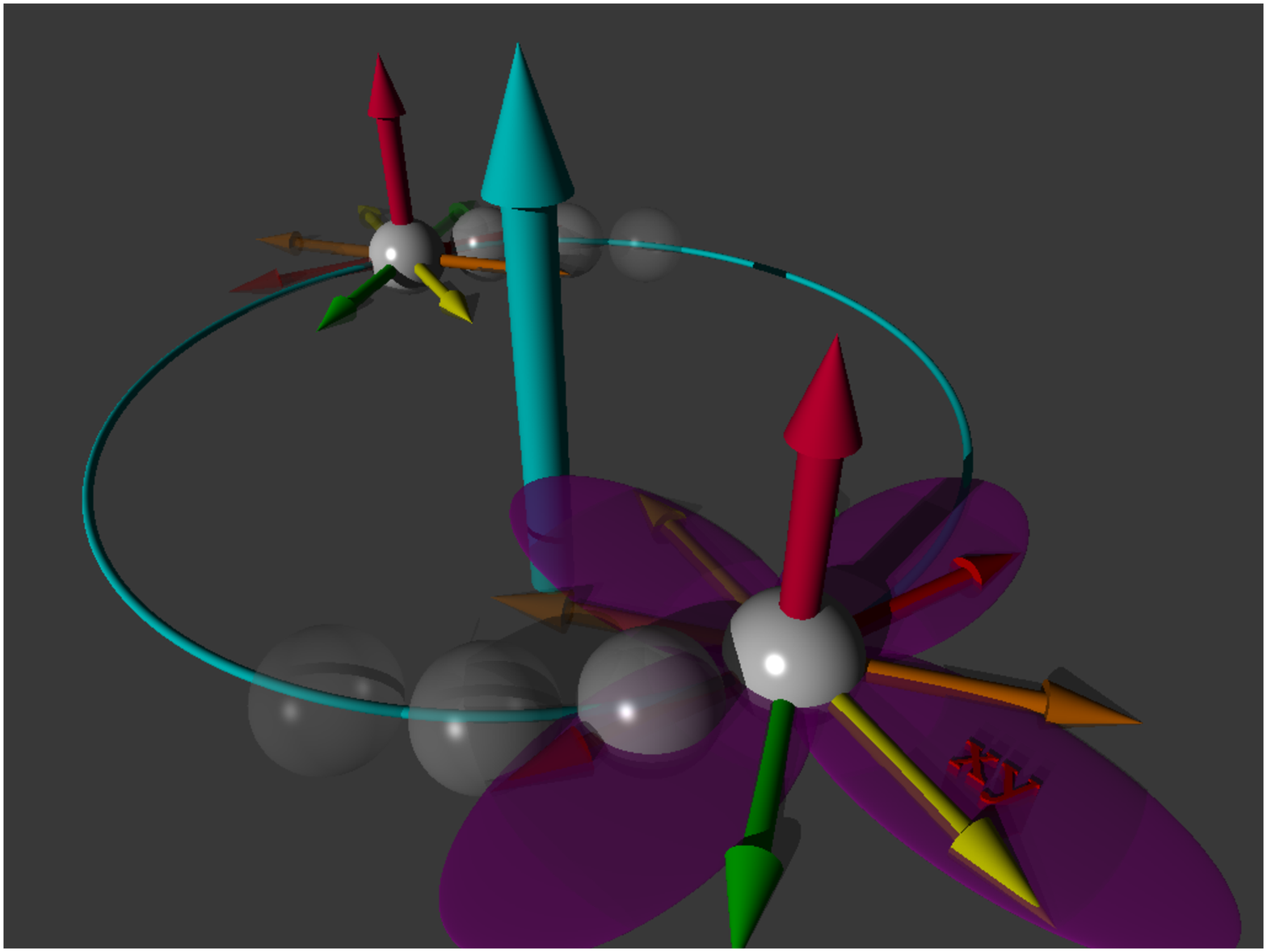}
        \put(3,70){\resizebox{70pt}{!}{{\color{green}(a)}}}
        \put(45,60){\resizebox{60pt}{!}{${\color{red}\vec{L}}$}}
        \put(22,60){\resizebox{45pt}{!}{${\color{red}\vec{d}}$}}
        \put(25,43){\resizebox{45pt}{!}{${\color{red}\vec{S}}$}}
      \end{overpic}} \\
    \resizebox{0.9\linewidth}{!}{
      \begin{overpic}{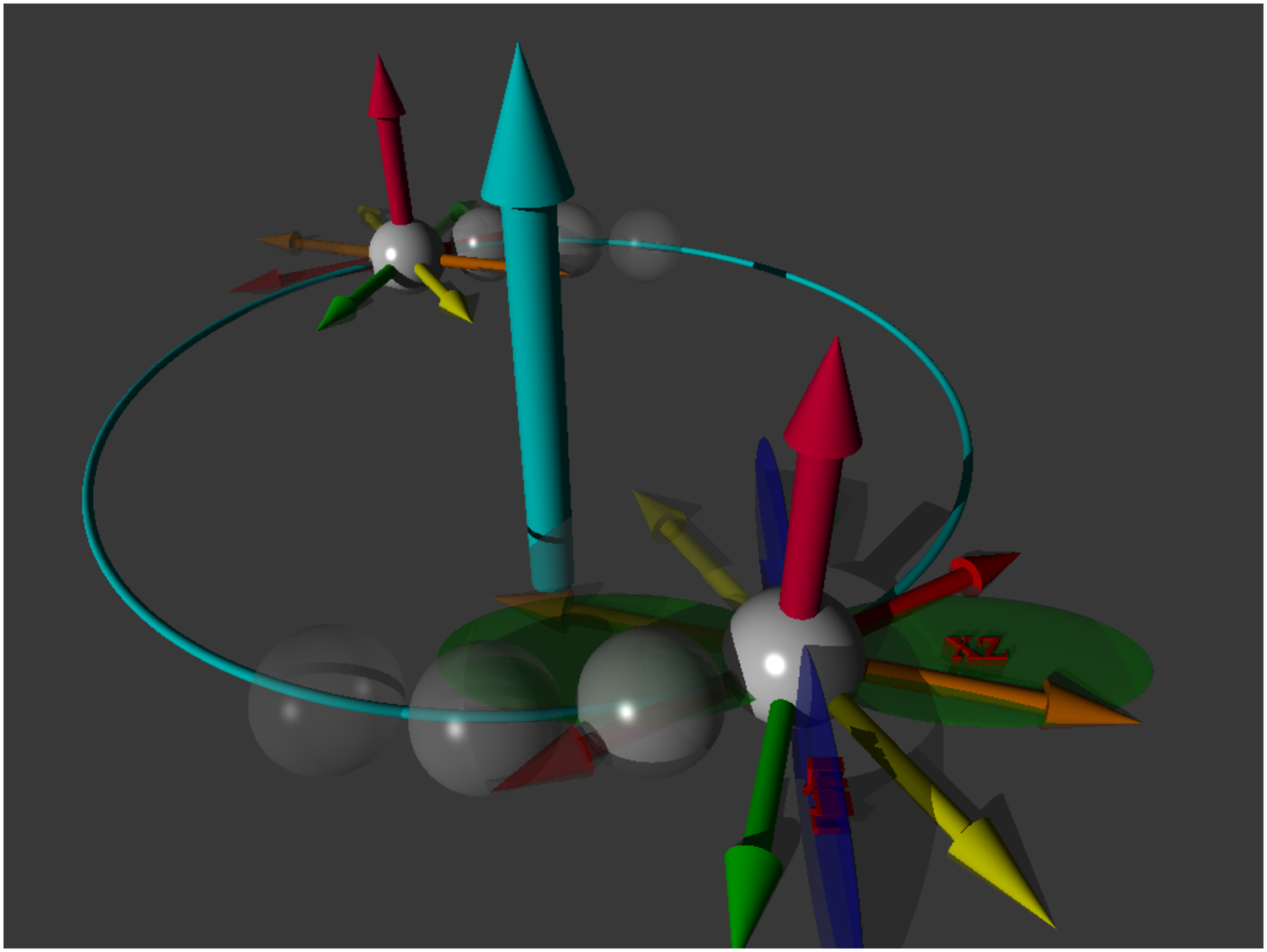}
        \put(3,70){\resizebox{70pt}{!}{{\color{green}(b)}}}
        \put(45,60){\resizebox{60pt}{!}{${\color{red}\vec{L}}$}}
        \put(22,60){\resizebox{45pt}{!}{${\color{red}\vec{d}}$}}
        \put(25,43){\resizebox{45pt}{!}{${\color{red}\vec{S}}$}}
      \end{overpic}}
  \end{tabular}
\caption{\label{cooper}(Color online) Schematic illustrations of a Cooper pair with spin $\vec{S}$ and orbital angular momentum $\vec{L}$ for a spin-triplet superconductor in the chiral state with $\vec{d}(\bm{k})$ formed in (a) the single band and (b) quasi-1D bands. In the former case, superconductivity is mainly from the $d_{xy}$ orbital, whereas the hybridization of $d_{xz}$ and $d_{yz}$ orbitals play a key role in the latter case. It is strongly desirable to distinguish these two scenarios.}
\end{figure}

In the single 2D band picture, a Cooper pair in the chiral $p$-wave state can be illustrated schematically in Fig.~\ref{cooper} (a), while its counterpart in the two quasi-1D band picture can be shown in Fig.~\ref{cooper} (b). It is highly demanding to distinguish the two possibilities by using available experimental probes. However, the fragile superconducting state of Sr$_2$RuO$_4$ does pose a challenge and restriction to experimental probes. For example, the low transition temperature with $T_c\approx 1.5$K is beyond the present technical limit of using the state-of-the-art angle-resolved photoemission spectroscopy~\cite{Damascelli2003}. In order to pin down this controversial issue with currently available experiments, in this work we study the vortex phase of a chiral $p$-wave superconductor for both the 2D band and quasi-1D band scenarios. It is found that different models give rise to qualitatively distinguishable vortex states, a promising characteristic to identify the underlying superconducting nature of Sr$_2$RuO$_4$.

The rest of this paper is organized as follows. In Sec.~\ref{sec:normal}, some basic properties of the Fermi surface in the normal state are studied. In Sec.~\ref{sec:super}, we construct the effective Hamiltonians based on the Bogoliubov-de Gennes theory for two different models as a pedagogical procedure. In Sec.~\ref{sec:result}, the results with focus on experimental observables are presented. We note that the vortex state for the single-band chiral $p$-wave model has been studied in previous literature~\cite{Takigawa2001}. In the present work we use more realistic parameters for the purpose of direct comparison of two different models. Finally, a concluding remark is given in Sec.~\ref{sec:conclusion}.

\section{Property of normal state: Fermi surface}\label{sec:normal}
Before proceeding to the discussion of different superconducting model, we briefly discuss the topology of the Fermi surface in the normal state. As we will see later, the Fermi surfaces of different bands give rise to different signature corresponding to the superconducting state. We begin with the Hamiltonian in the $4d-t_{2g}$ orbital basis  of the Ru ions
\begin{equation}\label{eqn:ham}
  \hat{H}_{\bm{k}}=\left(\begin{array}{ccc}
    \xi_1(\bm{k}) & g(\bm{k}) & 0\\
    g(\bm{k}) & \xi_2(\bm{k}) & 0 \\
    0 & 0 & \xi_3(\bm{k})
  \end{array}\right),
\end{equation}
where $\xi_1(\bm{k})\!\!=\!\!-2t\cos k_x-\mu$, $\xi_2(\bm{k})\!\!=\!\!-2t\cos k_y-\mu$, $g(\bm{k})\!=\!-4t'\sin k_x\sin k_y$ and $\xi_3(\bm{k})\!=\!-2t_3(\cos k_x+\cos k_y)-4t_3'\cos k_x\cos k_y-\mu_3$. Here 1, 2, and 3 denote orbitals $xz$, $yz$, and $xy$, respectively. Hereafter we take $(t,t',t_3,t_3')=(1,0.1,0.8,0.35)$ for the hopping parameters, and $(\mu,\mu_3)$ are fine tuned such that the electron density for each band is equal to $4/3$ \cite{Kontani2008}. Note that this set of parameters can reproduce Fermi surface, whose shape agrees with that obtained in the angle-resolved photoemission spectroscopy measurement above $T_c$ \cite{Damascelli2000}.

\begin{figure}[htb!]\centering
  \begin{tabular}{c}
    \resizebox{0.8\linewidth}{!}{
      \begin{overpic}{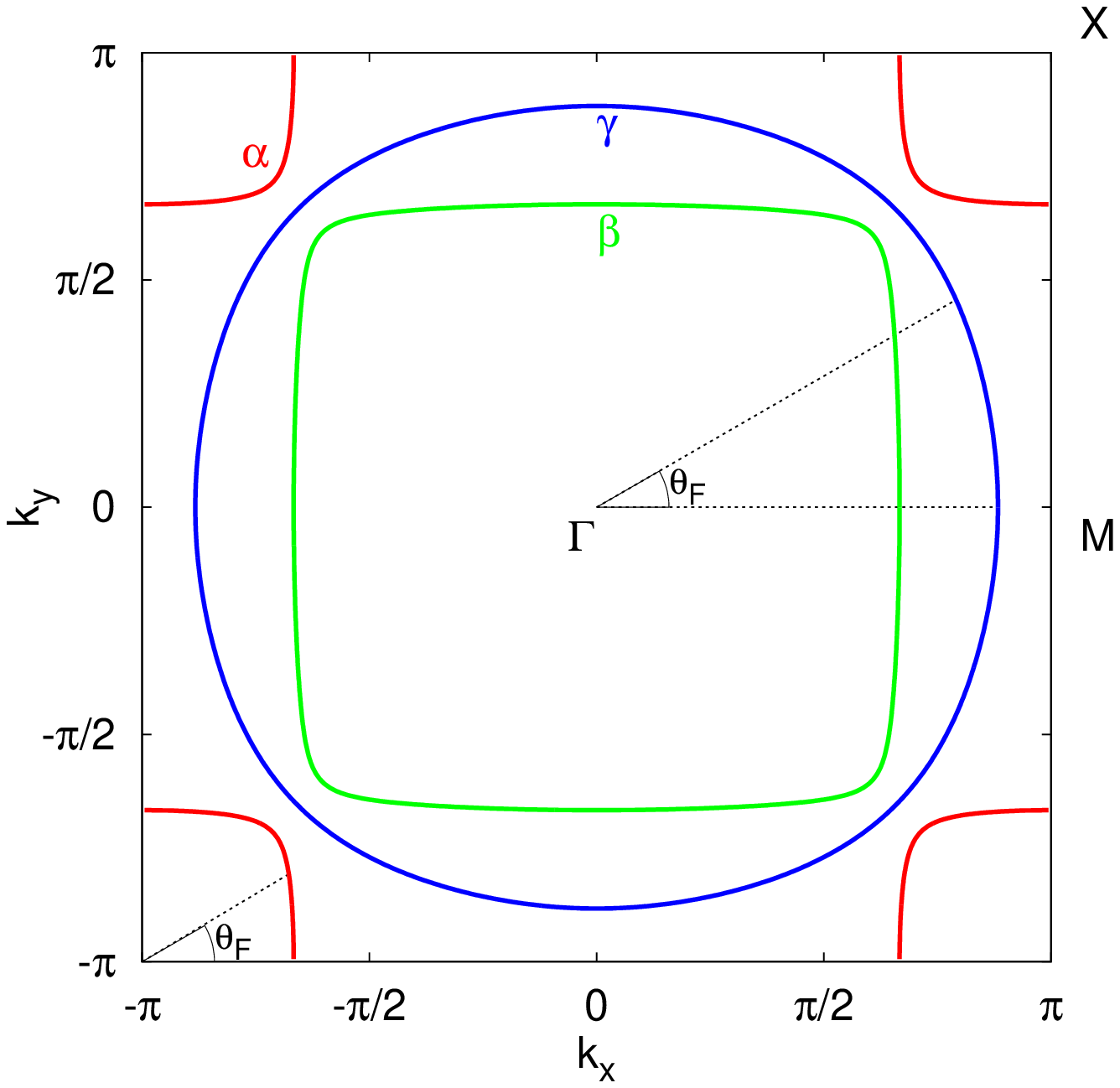}
        \put(0,85){\resizebox{25pt}{!}{(a)}}
      \end{overpic}} \\
    \resizebox{0.8\linewidth}{!}{
      \begin{overpic}{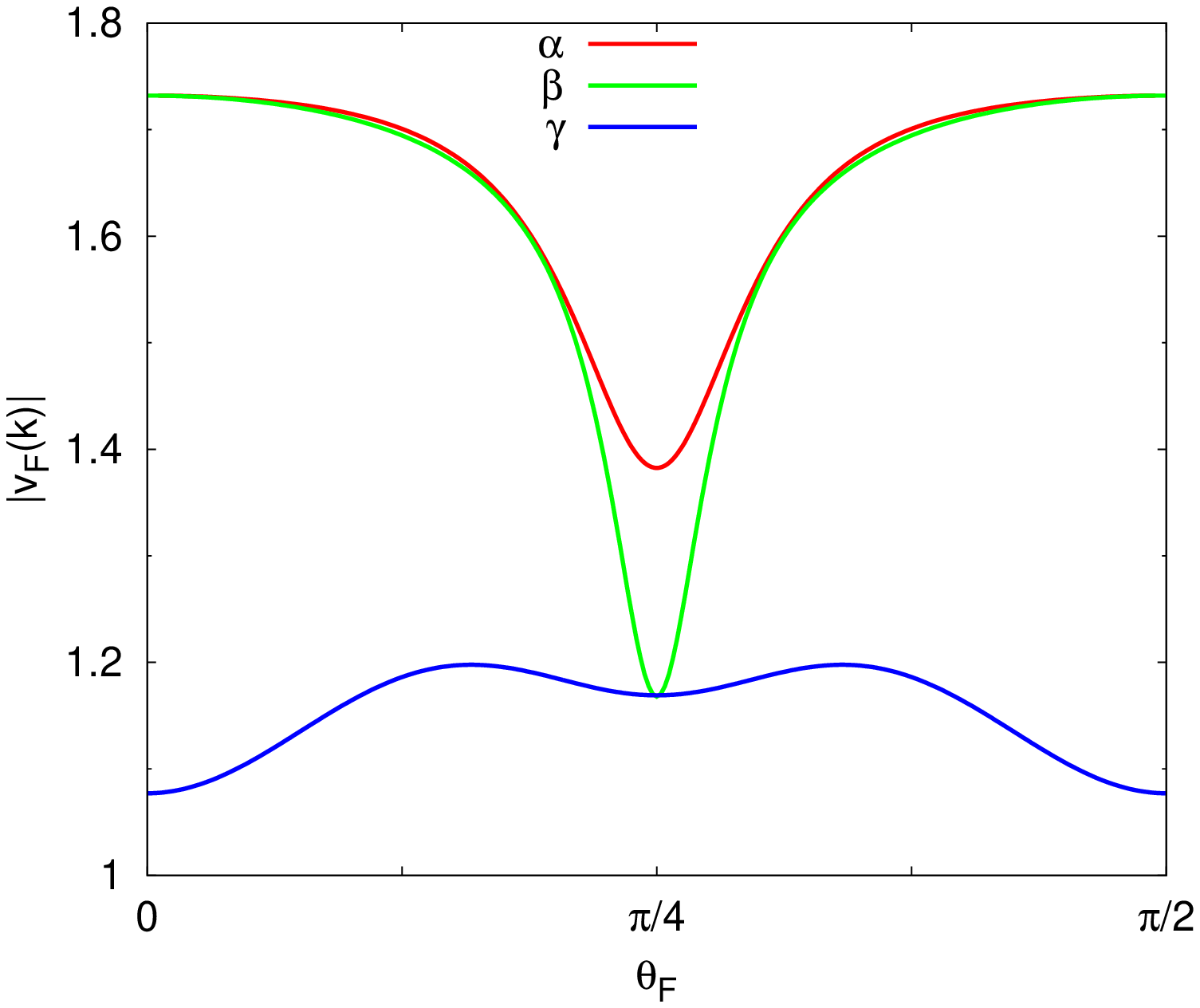}
        \put(0,85){\resizebox{25pt}{!}{(b)}}
      \end{overpic}}
  \end{tabular}
\caption{\label{pair}(Color online) (a) Two-dimensional Fermi surface for Sr$_2$RuO$_4$ in the normal state. (b) Fermi velocity $v_F(\bm{k})$ along the Fermi surface line. Here $\theta_F$ is the angle between the point $\bm{k}$ on the Fermi surface and $x$-axis, as indicated in the figure.}
\end{figure}

After diagonalizing the Hamiltonian in Eq.~\ref{eqn:ham}, the Fermi surface can be obtained, as shown in Fig.\ref{pair} (a). It is noted that the $\gamma$ band is almost isotropic in the two-dimensional plane, whereas the quasi-1D $\alpha,\beta$ bands show significant anisotropy. To further illustrate this point, we show the Fermi velocity $\vec{v}_F(\bm{k})$ in different bands in Fig.~\ref{pair} (b). We can see that the amplitude of $\vec{v}_F$ is maximized along the in-plane square lattice axis $a$ or $b$ for the $\alpha$ and $\beta$ bands, while it is almost isotropic for the $\gamma$ band.

In the subsequent sections related to the superconducting state, we will concentrate the discussion on the vortex state in order to provide a good test and the answer to the question in which band superconductivity takes place. As we will see later, the topology of Fermi surface is actually inherited in the superconducting state for which the band is predominant.

\section{Bogoliubov-de Gennes Theory of the Superconducting State}\label{sec:super}
\subsection{Single 2D band model}
First, we consider a single 2D band model based on the $d_{xy}$ orbital. In terms of the eigen-energy $E_\epsilon$ and the quasi-particle amplitudes $u^\epsilon_i$, $v^\epsilon_i$ at the $i$-th site, the Bogoliubov-de Gennes equations are given by\cite{Takigawa2001}
\begin{eqnarray}
\sum_j\left[\begin{array}{cc}
  H_{ij}&\Delta_{ij}\\
  \Delta^\dag_{ij}&-H^*_{ij}
\end{array}\right]\left[\begin{array}{c}
  u^\epsilon_j\\
  v^\epsilon_j
\end{array}\right]=E_\epsilon\left[\begin{array}{c}
  u^\epsilon_i\\
  v^\epsilon_i
\end{array}\right],
\end{eqnarray}
where $H_{ij}\!=\!-t_{ij}e^{i\varphi_{ij}}-\mu_3\delta_{i,j}$, and $\epsilon$ is an index of the eigenstate. The magnetic field is introduced through the Peierls phase factor $e^{i\varphi_{ij}}$ with $\varphi_{ij}=\frac{\pi}{\Phi_{0}}\int^{r_{i}}_{r_{j}}\mathbf{A(r)}\cdot d\mathbf{r}$, where $\bm{A}(\bm{r})=\frac{H}{2}(-y,x ,0)$ stands for the vector potential with magnetic field $H$ in the symmetric gauge and $\Phi_{0}=hc/2e$ is the superconducting flux quantum. Within this choice of gauge, the next-nearest vortices of the square-vortex lattice are located at the $45^\circ$ directions from the (100) direction. This vortex lattice configuration is suggested from the neutron-scattering experiment \cite{Riseman1998,Kealey2000,Takigawa2001}. We have $t_{ij}=t_3$ and $t_{ij}=t_3'$ for the nearest-neighbor and next-nearest-neighbor hopping, respectively.
The self-consistent equation for the pairing potential is reduced to
\begin{equation}
  \Delta_{ij}=\frac{V}{2}\delta_{i,j\pm\hat{e}}\sum_{\epsilon}(u^{\epsilon}_{j} v^{\epsilon\ast}_{i}-v^{\epsilon\ast}_{j}u^{\epsilon}_{i})\tanh\frac{E_{\epsilon}}{2T},
\end{equation}
with $T$ the temperature. Here $\hat{e}=\hat{x},\hat{y}$, denoting the unit vector along the $x$ and $y$ direction, respectively. In this paper, we set $V=1.0$.

The pair potential at each site $i$ can be decomposed into $p_x$ and $p_y$ components as
\begin{eqnarray}
  \Delta_{p_x}(\bm{r}_i)&=&\frac{\Delta_{\hat{x},i}-\Delta_{-\hat{x},i}}{2},\\
  \Delta_{p_y}(\bm{r}_i)&=&\frac{\Delta_{\hat{y},i}-\Delta_{-\hat{y},i}}{2}.
\end{eqnarray}
Here we have denoted
\begin{equation}
  \Delta_{\hat{e},i}=\Delta_{i,i+\hat{e}}\exp\left[i\frac{\pi}{\Phi_0}\int_{\bm{r}_i}^{\frac{\bm{r}_i+\bm{r}_{i+\hat{e}}}{2}}\bm{A}(\bm{r})\cdot\diff\bm{r}\right].
\end{equation}
For $\sin p_x\pm i\sin p_y$-wave superconductivity, we can define the paring potential as $\Delta_\pm(\bm{r}_i)\equiv\Delta_{p_x}(\bm{r}_i)\pm i\Delta_{p_y}(\bm{r}_i)$.

\subsection{Quasi-1D model}
We start with an effective two-orbital Hamiltonian that takes into account of only the Ru $d_{xz}$ and $d_{yz}$ orbitals \cite{Raghu2010a,Taylor2012,Imai2012}. By assuming an effective attraction that causes the $p$-wave superconducting pairing, one can construct an effective model to study the vortex physics of the chiral $p$-wave superconductors in the mixed state. Here we ignore the on-site repulsion, which is not expected to change our conclusions qualitatively in the present problem.

After the mean-field decomposition, one arrives at the Bogoliubov-de Gennes equations
\begin{multline}\label{eqn:bdg}
\sum_j\left(
\begin{array}{cccc}
H_{ij,mm} & \Delta_{ij,mm} & H_{ij,mn} & 0 \\
\Delta^\dag_{ij,mm} & -H^{\ast}_{ij,mm} & 0 & -H^{\ast}_{ij,mn} \\
H_{ij,mn} & 0 & H_{ij,nn} & \Delta_{ij,nn} \\
0 & -H^{\ast}_{ij,mn} & \Delta^\dag_{ij,nn} & -H^{\ast}_{ij,nn}
\end{array}
\right) \\
\times\left(
\begin{array}{cccc}
u^{\epsilon}_{j,m} \\
v^{\epsilon}_{j,m} \\
u^{\epsilon}_{j,n} \\
v^{\epsilon}_{j,n}
\end{array}
\right)= E_{\epsilon}\left(
\begin{array}{cccc}
u^{\epsilon}_{i,m} \\
v^{\epsilon}_{i,m} \\
u^{\epsilon}_{i,n} \\
v^{\epsilon}_{i,n}
\end{array}
\right),
\end{multline}
where $H_{ij,\alpha\beta}=-e^{i\varphi_{ij}}t_{ij,\alpha\beta}-\delta_{ij}\delta_{\alpha\beta}\mu$, and $u^{\epsilon}_{j,m}$, $u^{\epsilon}_{j,n}$, $v^{\epsilon}_{j,m}$ and $v^{\epsilon}_{j,n}$ are the Bogoliubov quasiparticle amplitudes on the $j$-th site with corresponding eigenvalues $E_{\epsilon}$. Here $m$ and $n$ denote the $d_{xz}$ and $d_{yz}$ orbitals, respectively. The hopping integrals are chosen as
\begin{equation}
t_{ij,\alpha\beta}=\begin{cases}
t&\text{$\alpha=\beta=m(n),i=j\pm\hat{x}(\hat{y})$},\\
-t'&\text{$\alpha\neq\beta,i=j\pm(\hat{x}+\hat{y})$},\\
t'&\text{$\alpha\neq\beta,i=j\pm(\hat{x}-\hat{y})$},\\
0&\text{otherwise}.
\end{cases}
\end{equation}

The matrix elements in the off-diagonal terms of Eq.~\ref{eqn:bdg} are obtained through the following self-consistent equations,
\begin{eqnarray}
\Delta_{ij,mm}=&&\frac{V}{2}\delta_{i,j\pm\hat{x}}\sum_{\epsilon}(u^{\epsilon}_{j,m} v^{\epsilon\ast}_{i,m}-v^{\epsilon\ast}_{j,m}u^{\epsilon}_{i,m})\tanh\frac{E_{\epsilon}}{2T},\nonumber\\
\Delta_{ij,nn}=&&\frac{V}{2}\delta_{i,j\pm\hat{y}}\sum_{\epsilon}(u^{\epsilon}_{j,n} v^{\epsilon\ast}_{i,n}-v^{\epsilon\ast}_{j,n}u^{\epsilon}_{i,n})\tanh\frac{E_{\epsilon}}{2T}.\nonumber \\
\end{eqnarray}

The orbital part of the order parameter can be decomposed as
\begin{eqnarray}
  \Delta_{p_x}(\bm{r}_i)&=&\frac{\Delta_{\hat{x},i}-\Delta_{-\hat{x},i}}{2},\\
  \Delta_{p_y}(\bm{r}_i)&=&\frac{\Delta_{\hat{y},i}-\Delta_{-\hat{y},i}}{2},
\end{eqnarray}
where we denote
\begin{eqnarray}
  \Delta_{\hat{x},i}&=&\Delta_{i,i+\hat{x},mm}\exp\left[i\frac{\pi}{\Phi_0}\int_{\bm{r}_i}^{\frac{\bm{r}_i+\bm{r}_{i+\hat{x}}}{2}}\bm{A}(\bm{r})\cdot\diff\bm{r}\right],\nonumber \\
  \Delta_{\hat{y},i}&=&\Delta_{i,i+\hat{y},nn}\exp\left[i\frac{\pi}{\Phi_0}\int_{\bm{r}_i}^{\frac{\bm{r}_i+\bm{r}_{i+\hat{y}}}{2}}\bm{A}(\bm{r})\cdot\diff\bm{r}\right].
\end{eqnarray}

For $\sin p_x\pm i\sin p_y$-wave superconductivity \footnote{It has been proposed that the order parameter in the quasi-1D model should assume a slightly different form, namely $\sin p_x\cos p_y\pm i\sin p_y\cos p_x$ \cite{Raghu2010a}. Nevertheless, we believe that most of our conclusions can be qualitatively applied to this case, too.}, we can define the paring potential as
\begin{equation}
  \Delta_\pm(\bm{r}_i)\equiv\Delta_{p_x}(\bm{r}_i)\pm i\Delta_{p_y}(\bm{r}_i).
\end{equation}
Without loss of generality, we restrict the calculations with $\Delta_+$ and $\Delta_-$ as the major and minor components of order parameters. Two types of vortices arise depending on the direction of the magnetic field insofar as the chirality is fixed \cite{Matsumoto2001}. In numerical computations, the unit cell with size $N_{x}\!\times\!N_{y}\!=\!33\!\times\!33$ and the number of such magnetic unit cells $M_{x}\!\times\!M_{y}\!=\!5\!\times\!5$ are used \cite{Wang1995}.

\subsection{Calculation of Experimental Observables}
In this subsection, we discuss the formalism for the local density of states and the nuclear spin-lattice relaxation rate in both models for Sr$_2$RuO$_4$.  In the scanning tunneling microscope (STM) experiment, the tunneling conductance is proportional to the local density-of-states (LDOS) $N(E,\mathbf{r}_{i})$, which can be calculated as
\begin{multline}
N(E,\mathbf{r}_{i})=-\sum_{\epsilon}[|\mathcal{U}_{i}^{\epsilon}|^{2}f'(E_{\epsilon}-E)+|\mathcal{V}_{i}^{\epsilon}|^{2} f'(E_{\epsilon}+E)].
\end{multline}
where $f'(E)$ is the derivative of the Fermi-Dirac distribution function with respect to energy. Hereafter we denote
\begin{equation}
  \mathcal{U}_i^\epsilon=u_i^\epsilon,\quad\mathcal{V}_i^\epsilon=v_i^\epsilon
\end{equation}
for the single-band model, and
\begin{equation}
  \mathcal{U}_i^\epsilon=u^{\epsilon}_{i,m}+u^{\epsilon}_{i,n},\quad\mathcal{V}_i^\epsilon=v^{\epsilon}_{i,m}+v^{\epsilon}_{i,n}
\end{equation}
for the quasi-1D model.

In addition to the STM measurement, nuclear magnetic resonance (NMR) is another related powerful method to identify distinct signatures predicted by different models. Generally speaking, this method is able to simultaneously shed light on the spatial profile of the zero-energy quasi-particles through the relaxation time $T_1$. The nuclear spin-lattice relaxation rate we consider is given by~\cite{Takigawa1999,Jiang2011}

\begin{eqnarray}\label{eqn:nmr}
R(r_{i},r_{i'})=&&\textmd{Im}\chi_{+,-}(r_{i},r_{i'},i\Omega_{n}\rightarrow\Omega+i\eta)/(\Omega/T)|_{\Omega\rightarrow 0}\nonumber\\
=&&-\sum_{\epsilon,\epsilon'}\mathcal{U}^{\epsilon}_{i}\mathcal{U}^{\epsilon'\ast}_{i}[\mathcal{U}^{\epsilon}_{i'}\mathcal{U}^{\epsilon'\ast}_{i'}+\mathcal{V}^{\epsilon}_{i'}\mathcal{V}^{\epsilon'\ast}_{i'}]\nonumber\\
&&\times\pi T f'(E_{\epsilon})\delta(E_{\epsilon}-E_{\epsilon'}).
\end{eqnarray}
We choose $\textbf{r}_{i}=\textbf{r}_{i'}$ by considering that the nuclear spin-lattice relaxation at a local site is dominant. Then the site-dependent relaxation time is given by $T_{1}(r)\!=\!1/R(r,r)$. Roughly speaking, $T_1$ is proportional to the integral of the LDOS within the energy range $0\!\leq\!E\!\leq\!T$. Therefore, the NMR experiment is also expected to provide important fingerprints for the two models.

\section{Results}\label{sec:result}
In this section, we will present results for the vortex states based on the two different models and basic methods discussed in the previous section.
\subsection{Vortex Structure}
To begin with, a general picture on the vortex structure for the superconducting order parameters for both models will be shown below. We will also show the spatial dependence of the pairing order parameter and LDOS at zero bias. While the former cannot be measured directly, the LDOS is an experimentally observable quantity in the STM measurement.

\begin{figure}[ht]\centering
  \begin{tabular}{cc}
    \resizebox{0.5\linewidth}{!}{
      \begin{overpic}{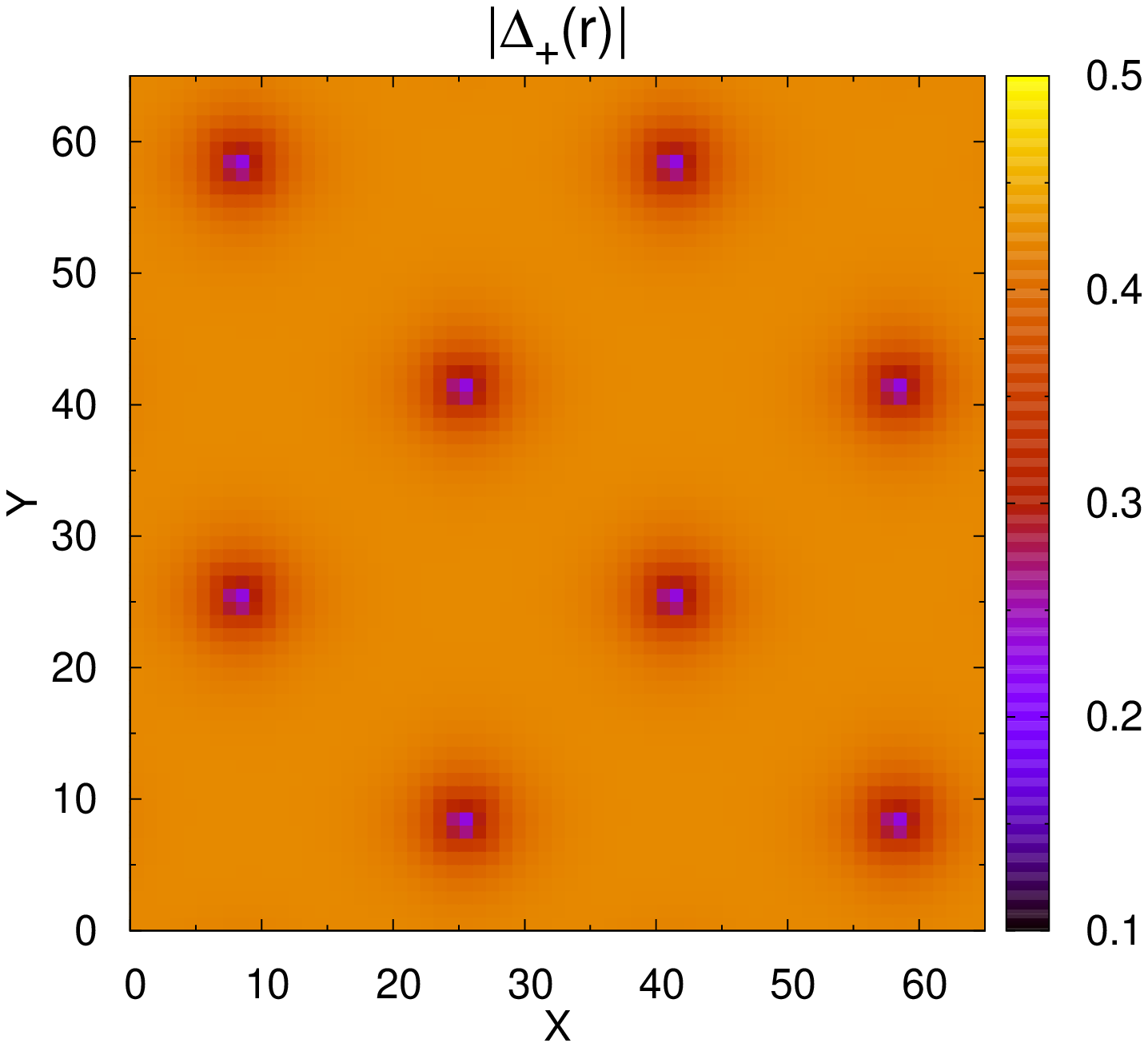}
        \put(10,80){\resizebox{25pt}{!}{(a)}}
      \end{overpic}} &
    \resizebox{0.5\linewidth}{!}{
      \begin{overpic}{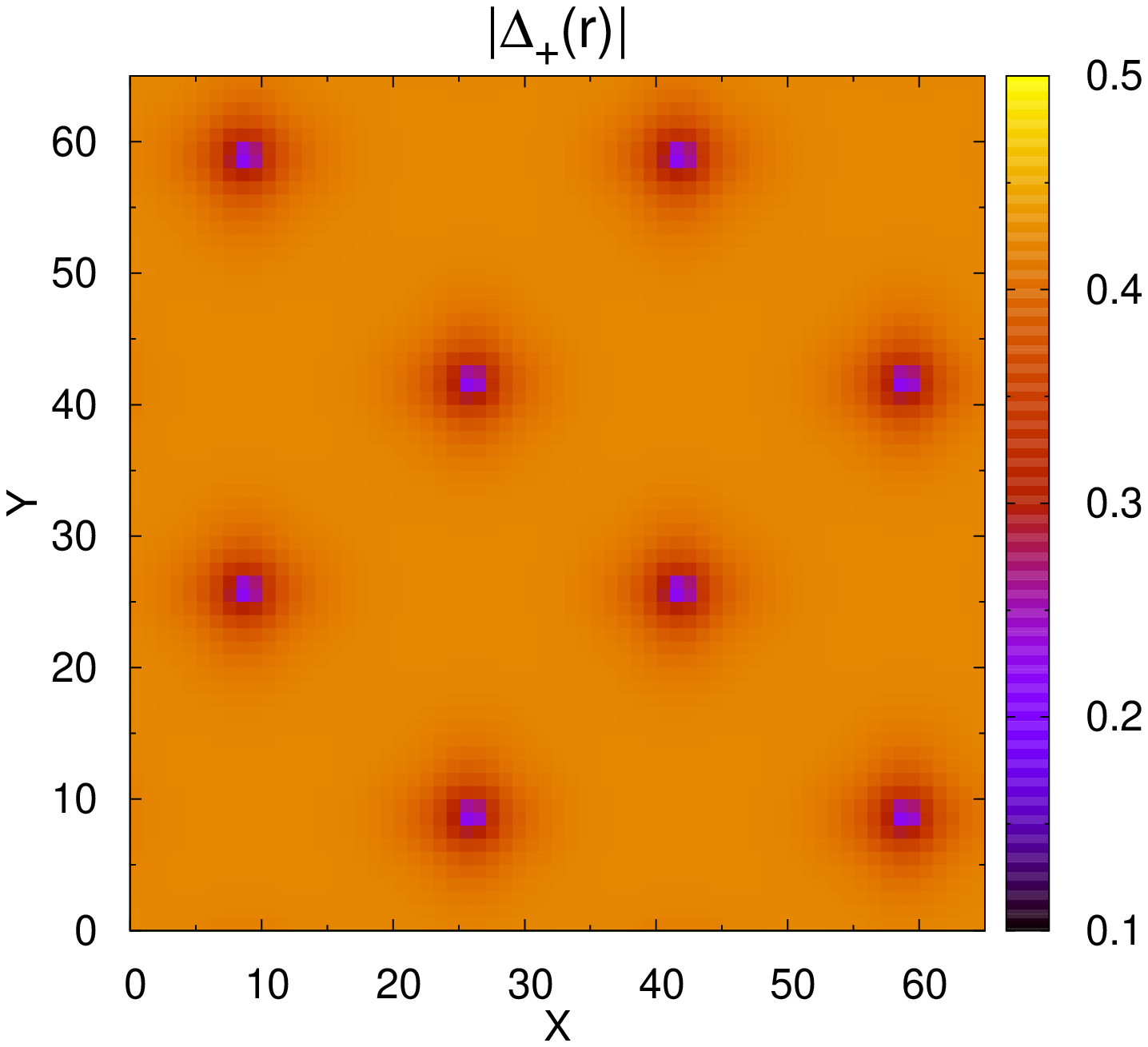}
        \put(10,80){\resizebox{25pt}{!}{(b)}}
      \end{overpic}} \\
    \resizebox{0.5\linewidth}{!}{
      \begin{overpic}{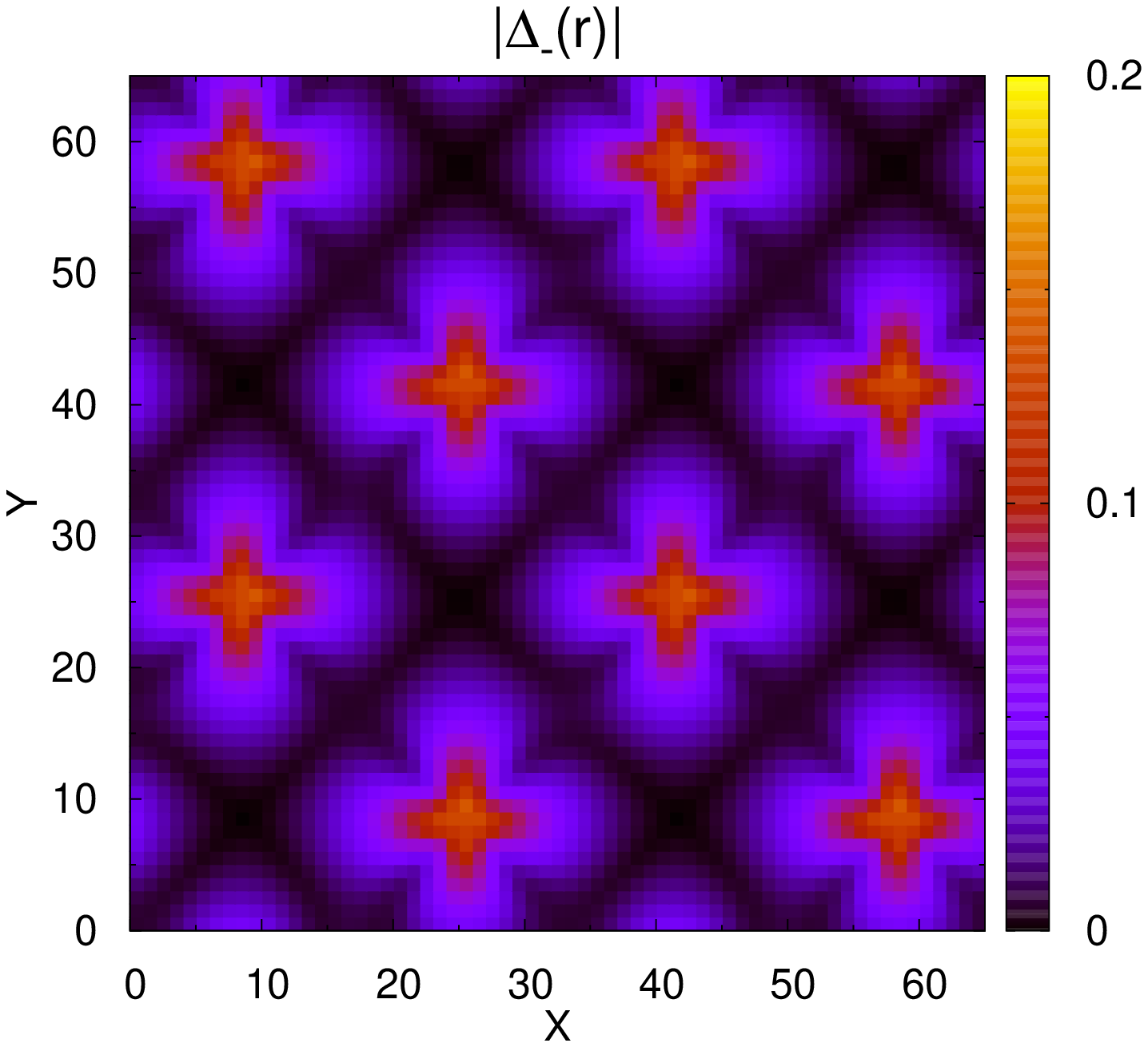}
        \put(10,80){\resizebox{25pt}{!}{(c)}}
      \end{overpic}} &
    \resizebox{0.5\linewidth}{!}{
      \begin{overpic}{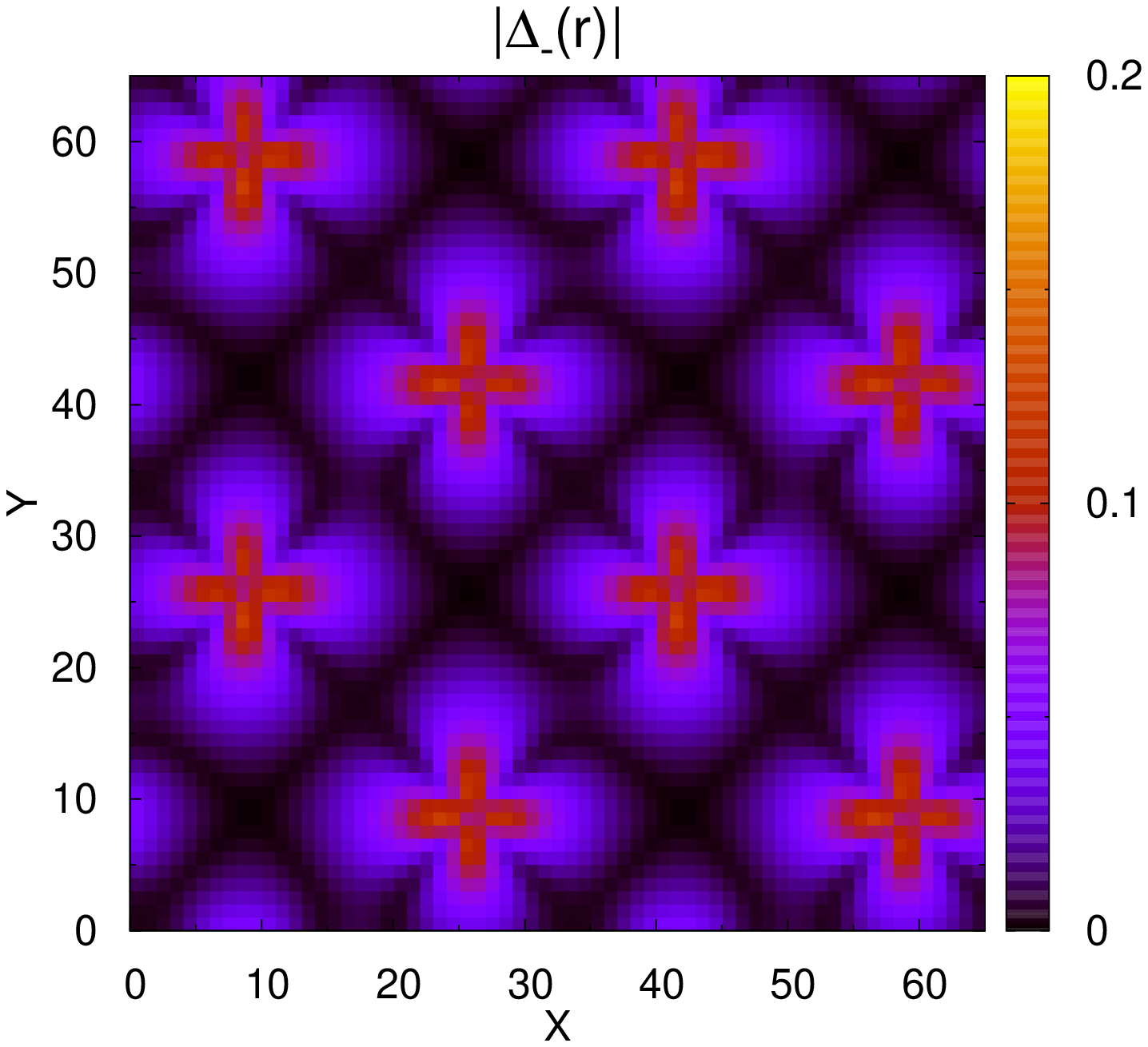}
        \put(10,80){\resizebox{25pt}{!}{(d)}}
      \end{overpic}} \\
    \resizebox{0.5\linewidth}{!}{
      \begin{overpic}{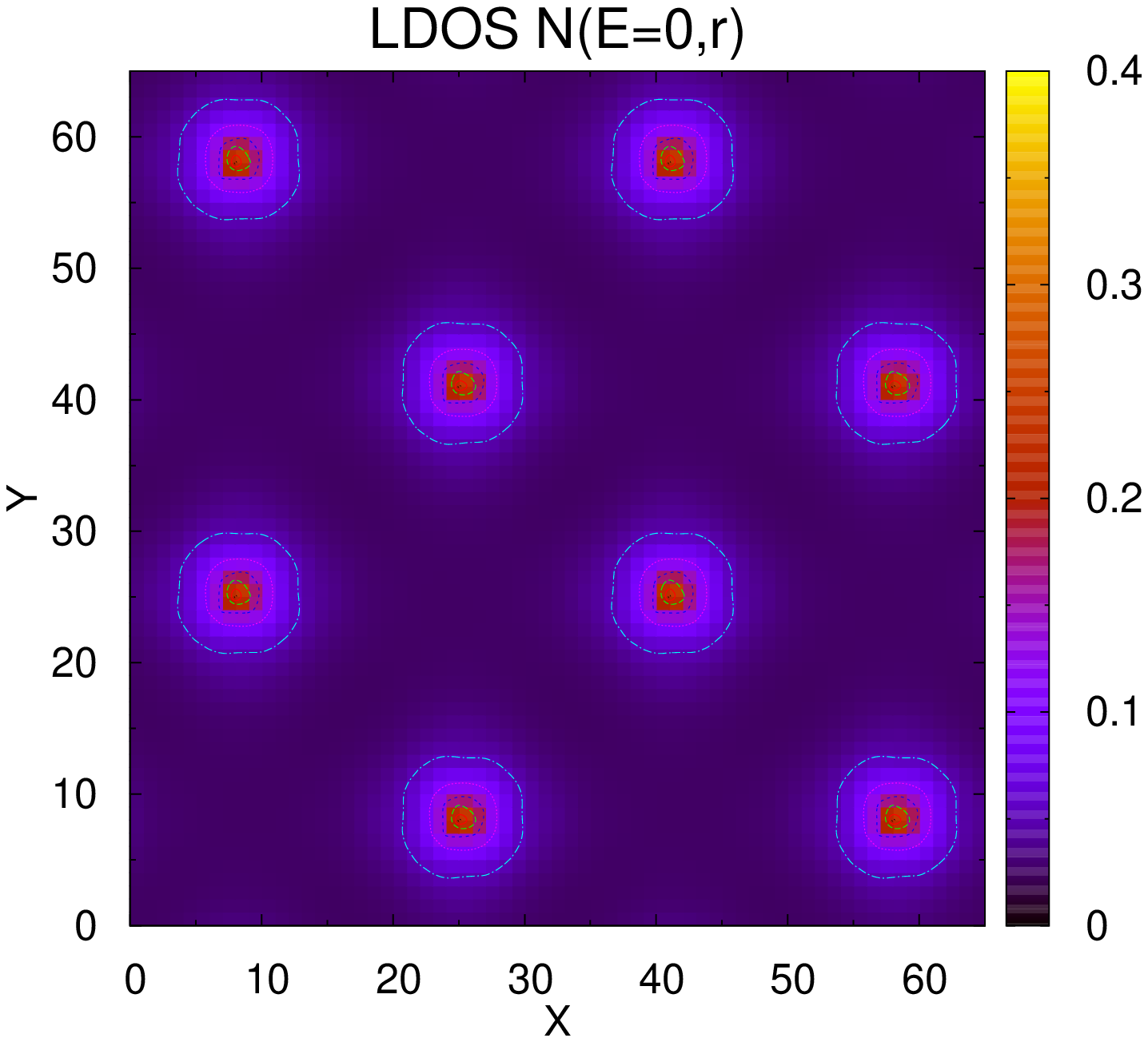}
        \put(10,80){\resizebox{25pt}{!}{(e)}}
      \end{overpic}} &
    \resizebox{0.5\linewidth}{!}{
      \begin{overpic}{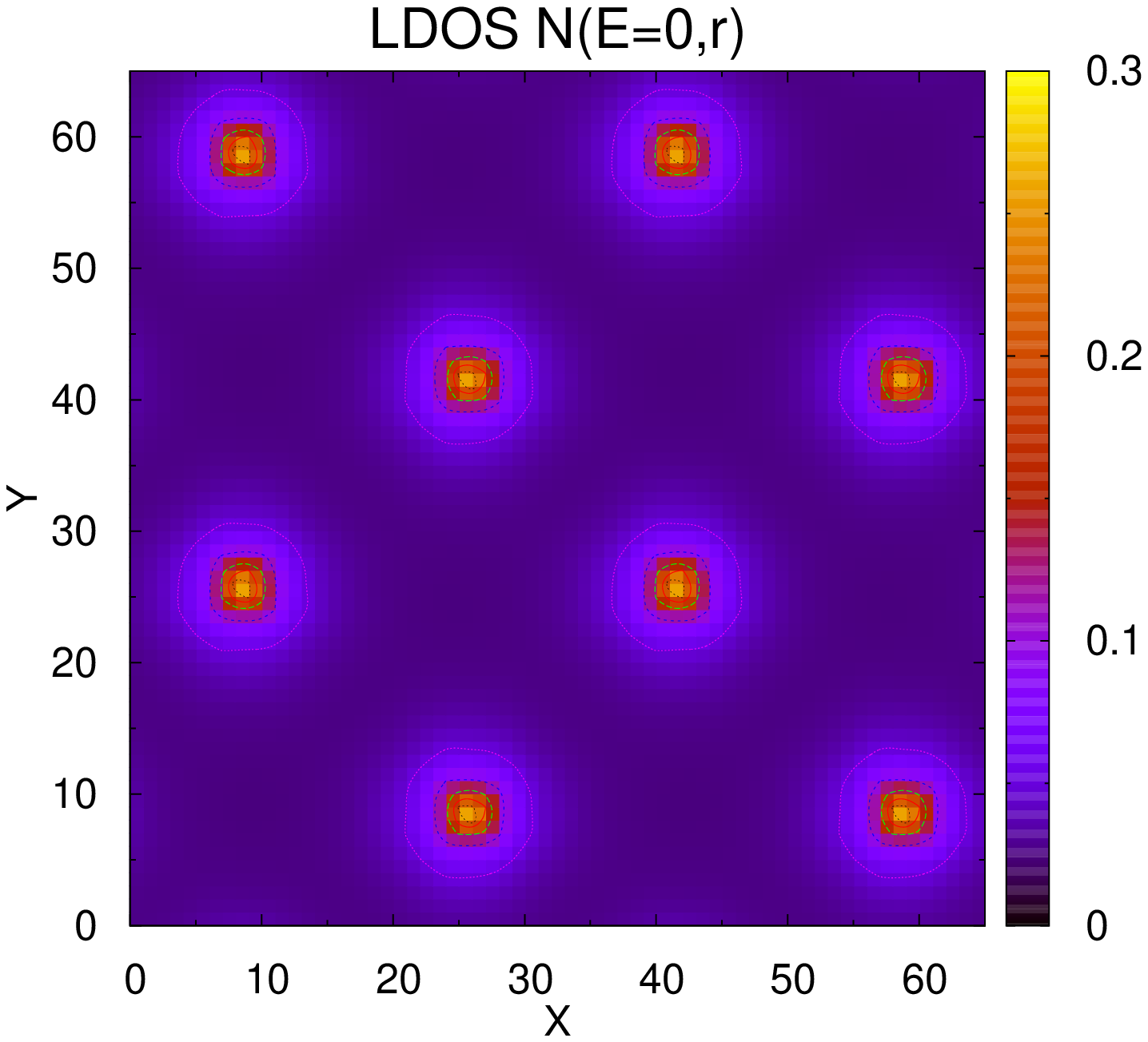}
        \put(10,80){\resizebox{25pt}{!}{(f)}}
      \end{overpic}}
  \end{tabular}
\caption{\label{single}(Color online) Vortex lattice structure in a single band model. Left panels are for a negative vortex lattice, and right panels for a positive vortex lattice. Upper panels (a) and (b): Spatial distribution of the major component $|\Delta_+(\bm{r})|$; Middle panels (c) and (d): Spatial distribution of the admixed component $|\Delta_-(\bm{r})|$; Lower panels (e) and (f): LDOS $N(E\!=\!0,\bm{r})$. The size of a magnetic unit cell is $33\times33$.}
\end{figure}

First, we study the single band model and plot the order parameters as a function of position in a vortex lattice in Fig.~\ref{single}. Due to the broken time-reversal symmetry of the chiral $p$-wave state, there are two types of vortices depending on the direction of the magnetic field. However, the negative vortex with winding number opposite to the chirality is the stable state \cite{Heeb1999}. And as shown in Fig.~\ref{single}, the orientation of the square shape is different depending on the winding. In the negative or positive vortex case, the shape of $|\Delta_+|$ around the vortex core is nearly isotropic with minor anisotropy. However, the induced component $|\Delta_-|$ in both negative and positive vortex shows similar shapes, which extends along the $a$-axis. Finally, as shown in Fig.~\ref{single}(c) and (f) we also calculate the LDOS $N(E=0,\bm{r})$ which is related to the tunneling conductance at zero bias in the STM experiment. We can see that the LDOS comes to a peak around the vortex core, with slightly anisotropic extension along the (110) direction. This is because that the amplitude of Fermi velocity $v_F$ for the $\gamma$ band is almost isotropic with slight enhancement near the (110) direction, as shown in Fig.~\ref{pair} (b). Such features are qualitatively similar to previous results~\cite{Takigawa2001}.

\begin{figure}[ht]\centering
  \begin{tabular}{cc}
    \resizebox{0.5\linewidth}{!}{
      \begin{overpic}{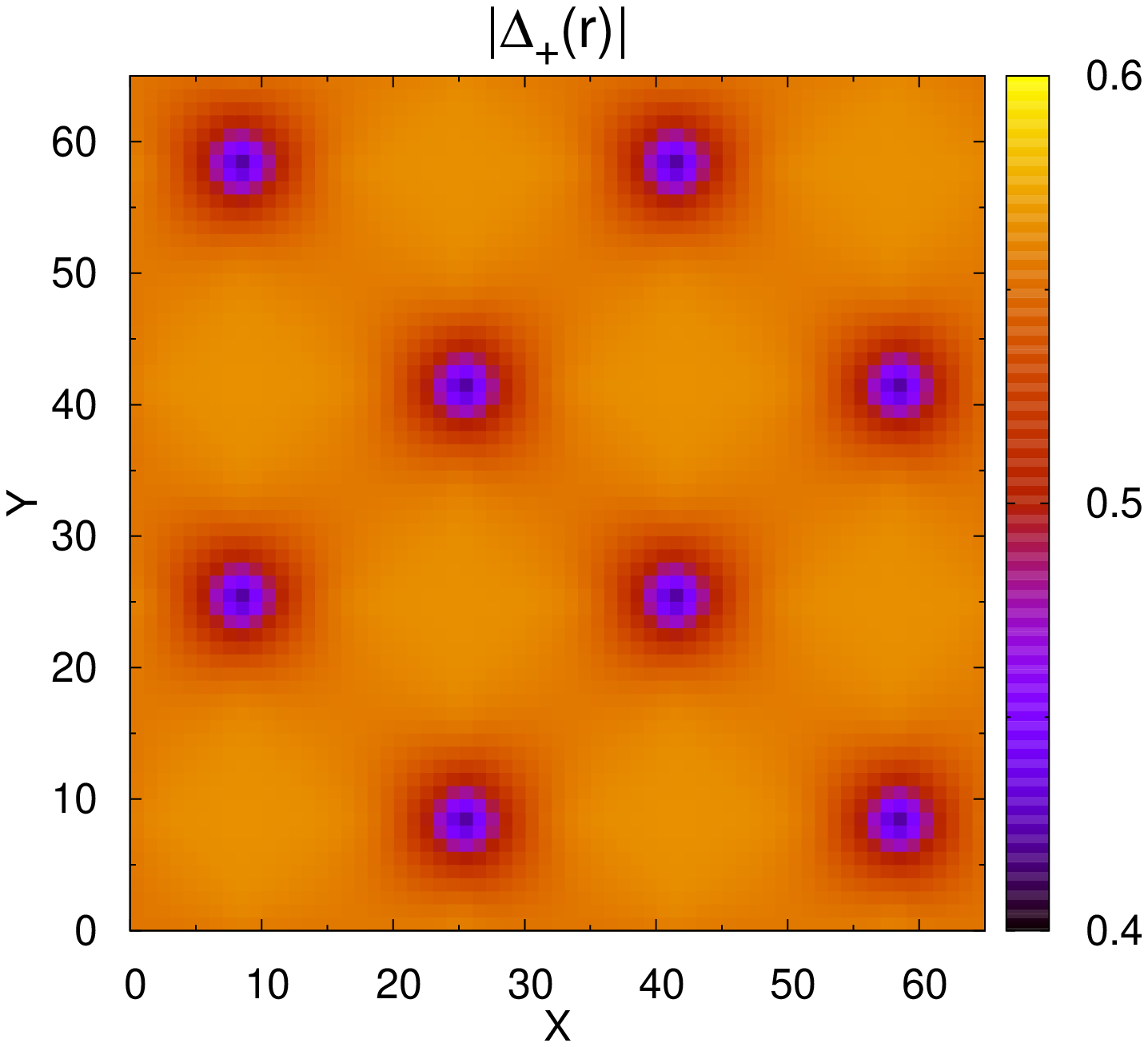}
        \put(10,80){\resizebox{25pt}{!}{(a)}}
      \end{overpic}} &
    \resizebox{0.5\linewidth}{!}{
      \begin{overpic}{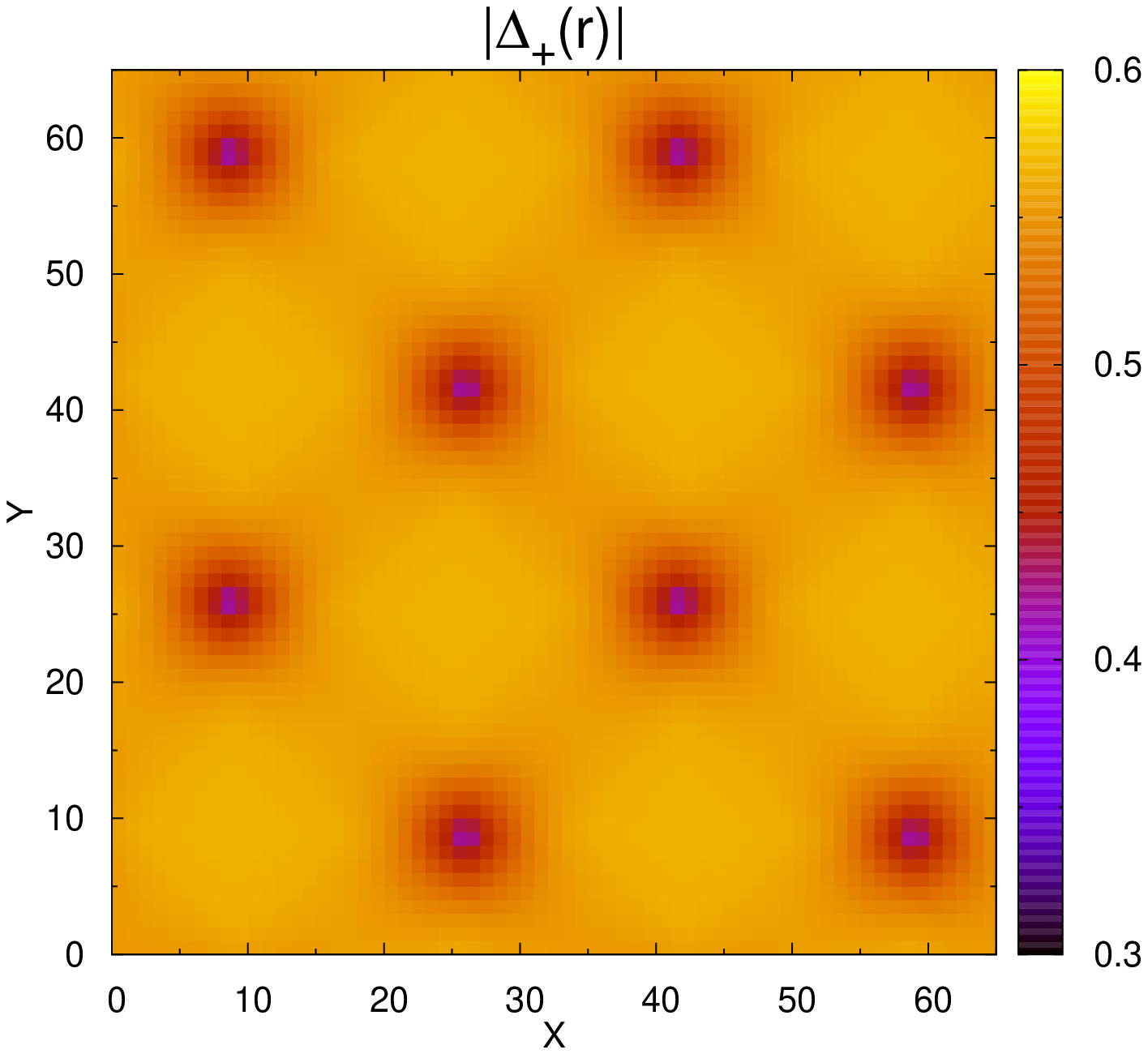}
        \put(10,80){\resizebox{25pt}{!}{(b)}}
      \end{overpic}} \\
    \resizebox{0.5\linewidth}{!}{
      \begin{overpic}{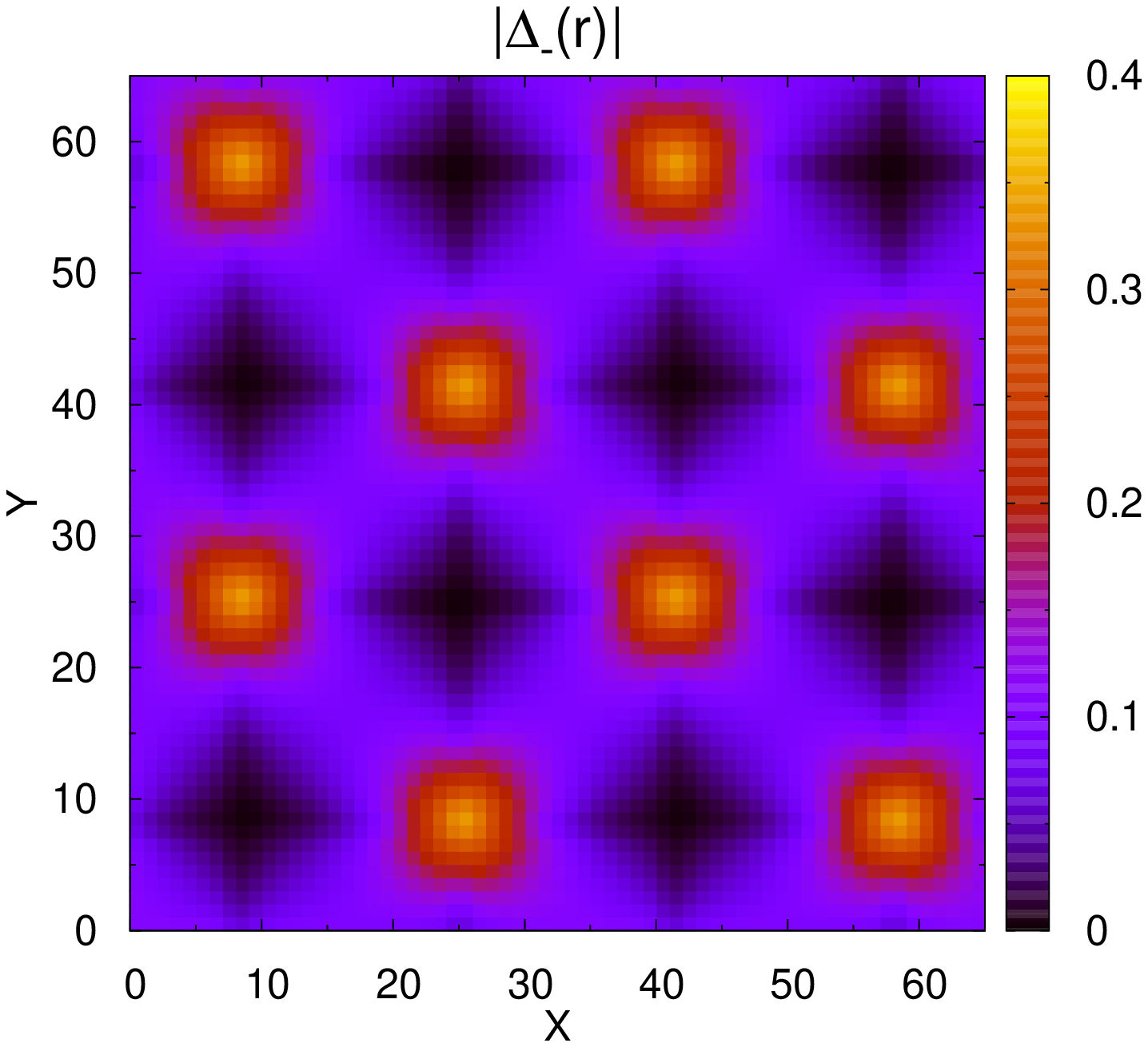}
        \put(10,80){\resizebox{25pt}{!}{(c)}}
      \end{overpic}} &
    \resizebox{0.5\linewidth}{!}{
      \begin{overpic}{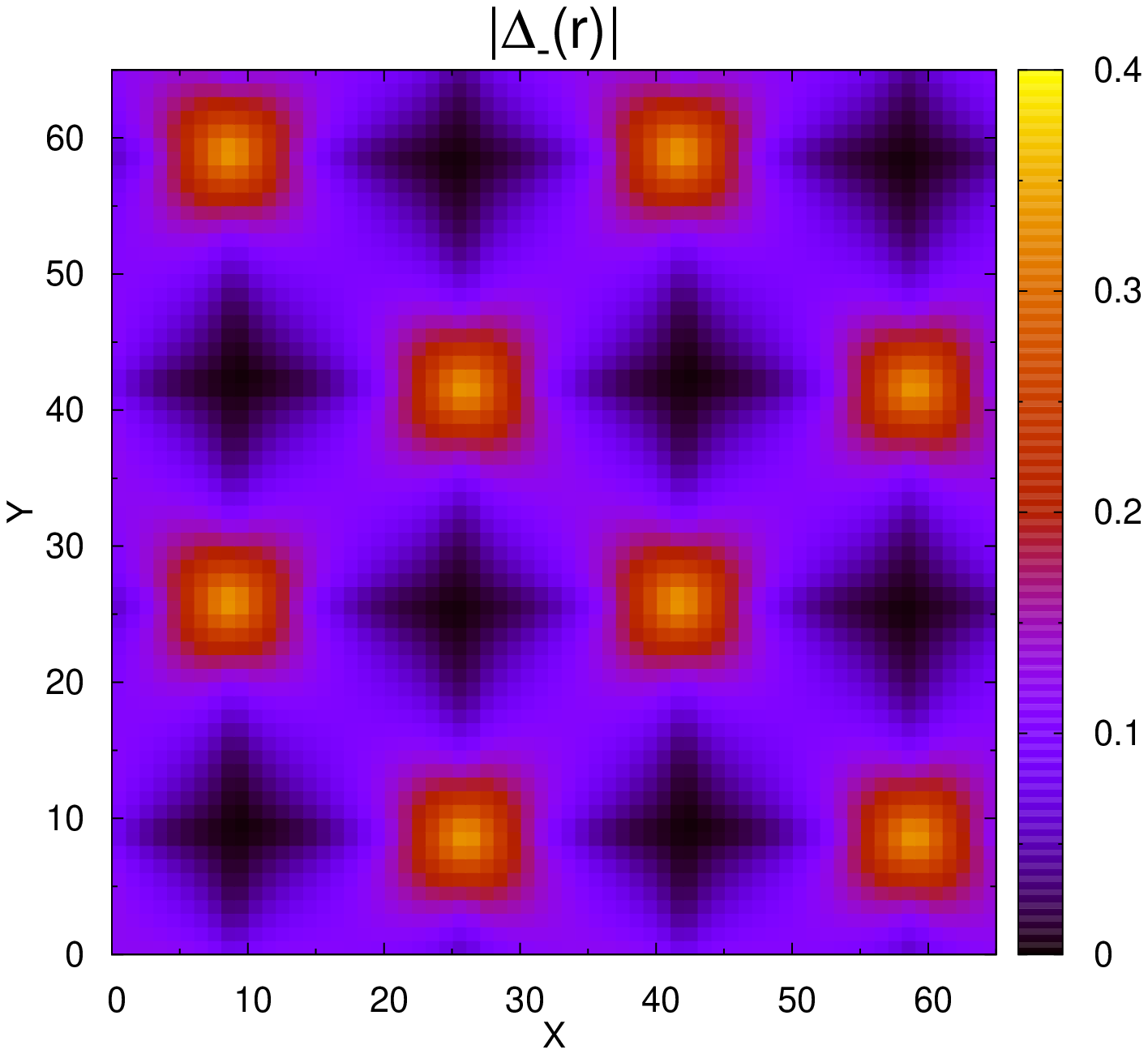}
        \put(10,80){\resizebox{25pt}{!}{(d)}}
      \end{overpic}} \\
    \resizebox{0.5\linewidth}{!}{
      \begin{overpic}{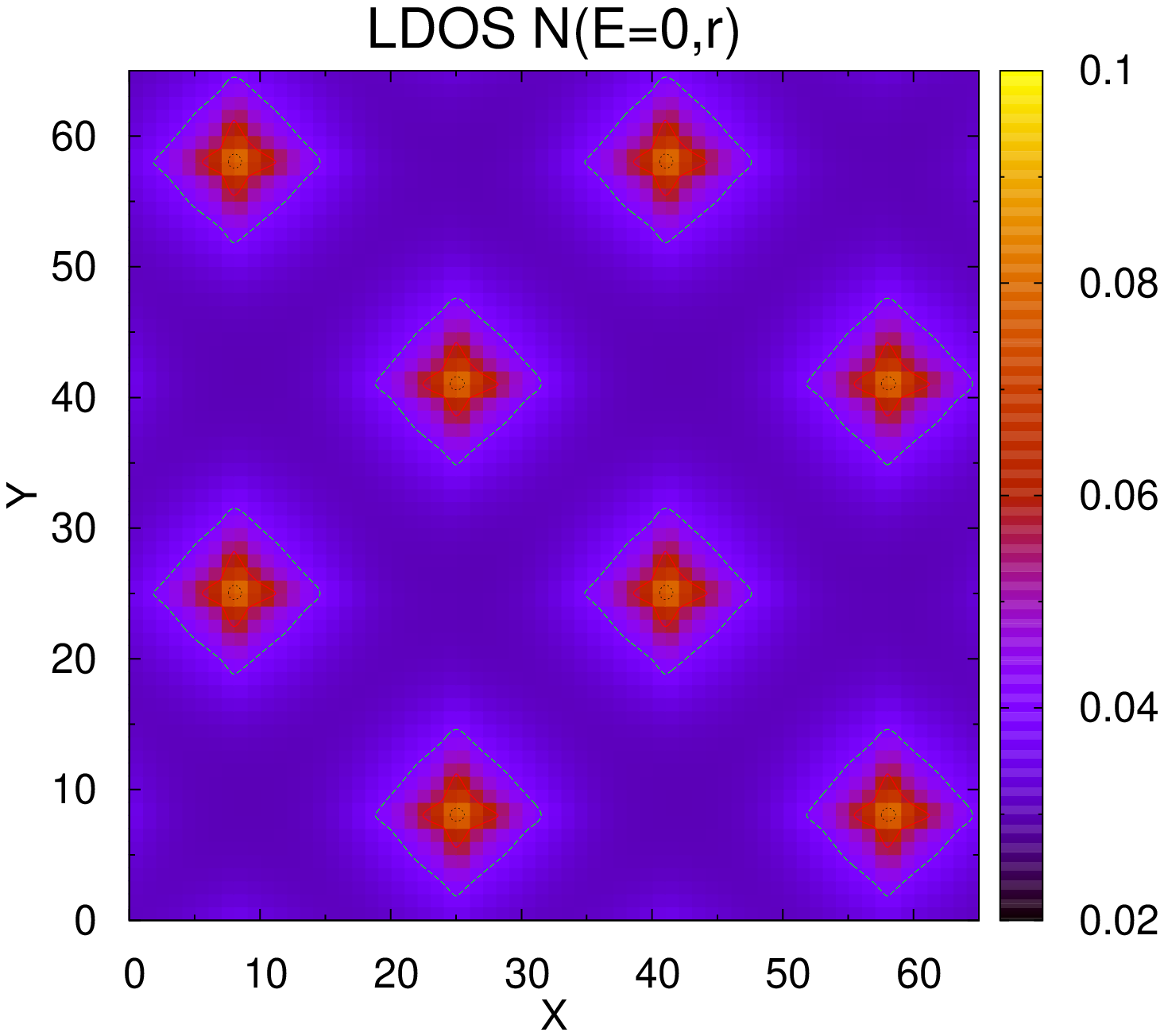}
        \put(10,80){\resizebox{25pt}{!}{(e)}}
      \end{overpic}} &
    \resizebox{0.5\linewidth}{!}{
      \begin{overpic}{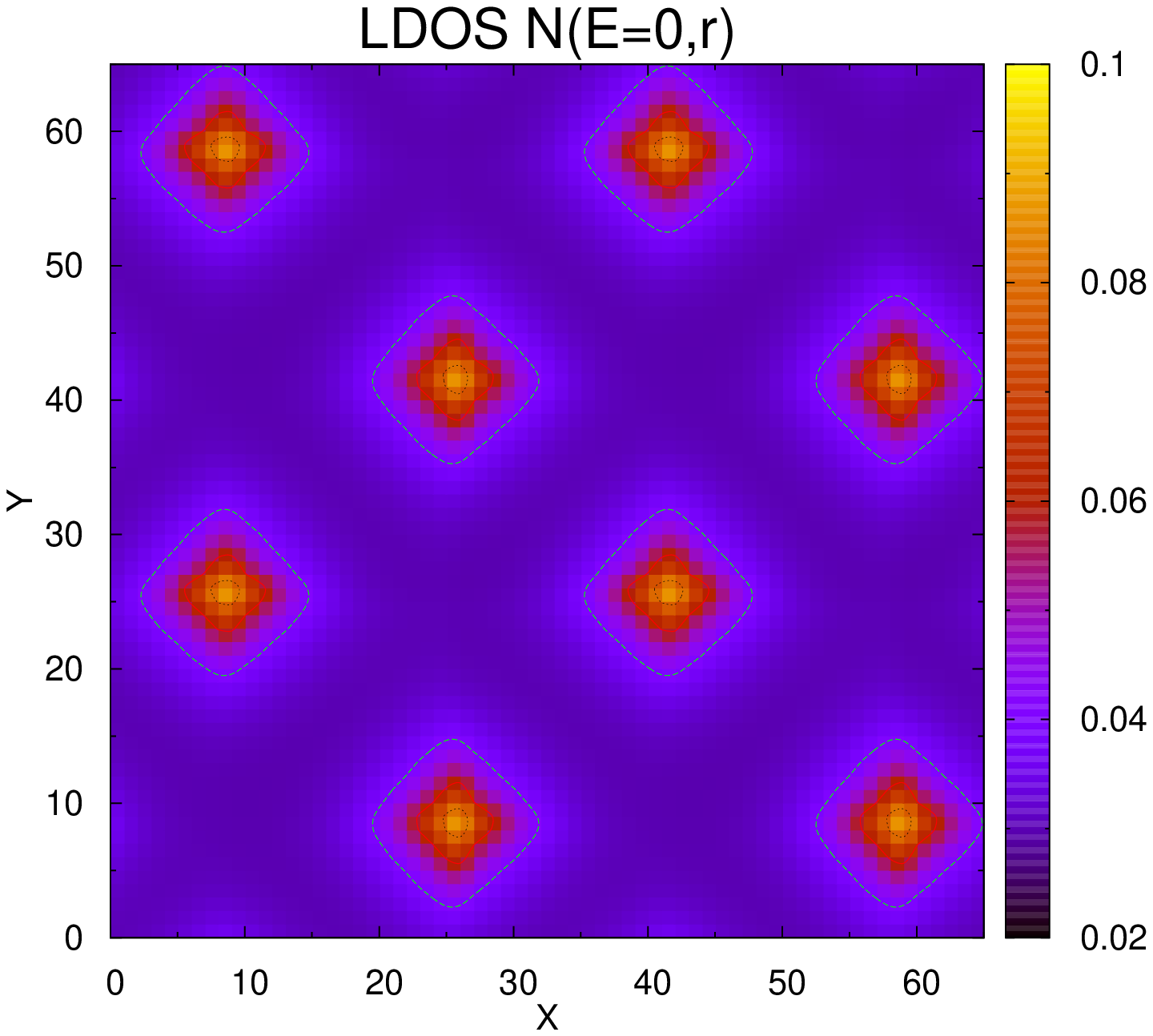}
        \put(10,80){\resizebox{25pt}{!}{(f)}}
      \end{overpic}}
  \end{tabular}
\caption{\label{multi}(Color online) Vortex lattice structure in a two band model. Left panels are for a negative vortex lattice, and right panels for a positive vortex lattice. Upper panels (a) and (b): Spatial distribution of the major component $|\Delta_+(\bm{r})|$; Middle panels (c) and (d): Spatial distribution of the admixed component $|\Delta_-(\bm{r})|$; Lower panels (e) and (f): LDOS at zero bias $N(E\!=\!0,\bm{r})$. The size of a magnetic unit cell is $33\times33$.}
\end{figure}

Now we turn to the results for the quasi-1D model. Similar calculations as the single band case are performed with the corresponding results shown in Fig.~\ref{multi}. We find several qualitative differences resulting from a different models after careful comparison. First, in this model, the amplitude of the major component $\Delta_+(\bm{r})$ is not substantially suppressed at the vortex core, compared with the previous model. This can be seen from the density scale of Fig.~\ref{multi} (a) and (b) showing a variation from about 0.3 to 0.6, whereas the counterpart for the single band model in Fig.~\ref{single} (a) and (b) is from 0.1 to 0.5. Secondly, for the induced component $\Delta_-(\bm{r})$, its amplitude reaches a maximum at the vortex core. On the contrary, $|\Delta_-(\bm{r})|$ maximizes near the core along the $a$ or $b$ axis and shows a fourfold symmetry.

More importantly, one should pay attention to the LDOS at zero bias for this model. At a first glance, the shape of LDOS around the vortex resembles a rhombus (\greendiamond), with its vertices pointing along the $a$ or $b$ axis. Such a highly anisotropic structure is actually an indication of strong anisotropy for the angle-resolved Fermi velocity. This intuitive understanding is supported by our calculations showing that $|\vec{v}_F|$ in the normal state is maximized along the $a$ or $b$ axis for both $\alpha$ and $\beta$ bands in Fig.~\ref{pair} (b). As a result, the shape of the LDOS at zero bias, acting as a fingerprint of the active band where the superconductivity arises, can be used to distinguish the two different models in the STM measurement.

Moreover, we find that the peak feature at the vortex core is substantially smeared for the quasi-1D model (pay attention to the density scale of Fig.~\ref{multi} (e) and (f) in comparison with the single-band counterpart in Fig.~\ref{single} (e) and (f)). This property is unique for a quasi-1D model \cite{Takigawa2006}, and the underlying physics is the strong suppression of low-energy resonance in the vortex state of a quasi-1D model, which will be discussed in detail with more experimental signatures below.

\subsection{Signatures in STM and NMR Measurements}
\begin{figure}[htb!]\centering
  \begin{tabular}{c}
    \resizebox{0.8\linewidth}{!}{
      \begin{overpic}{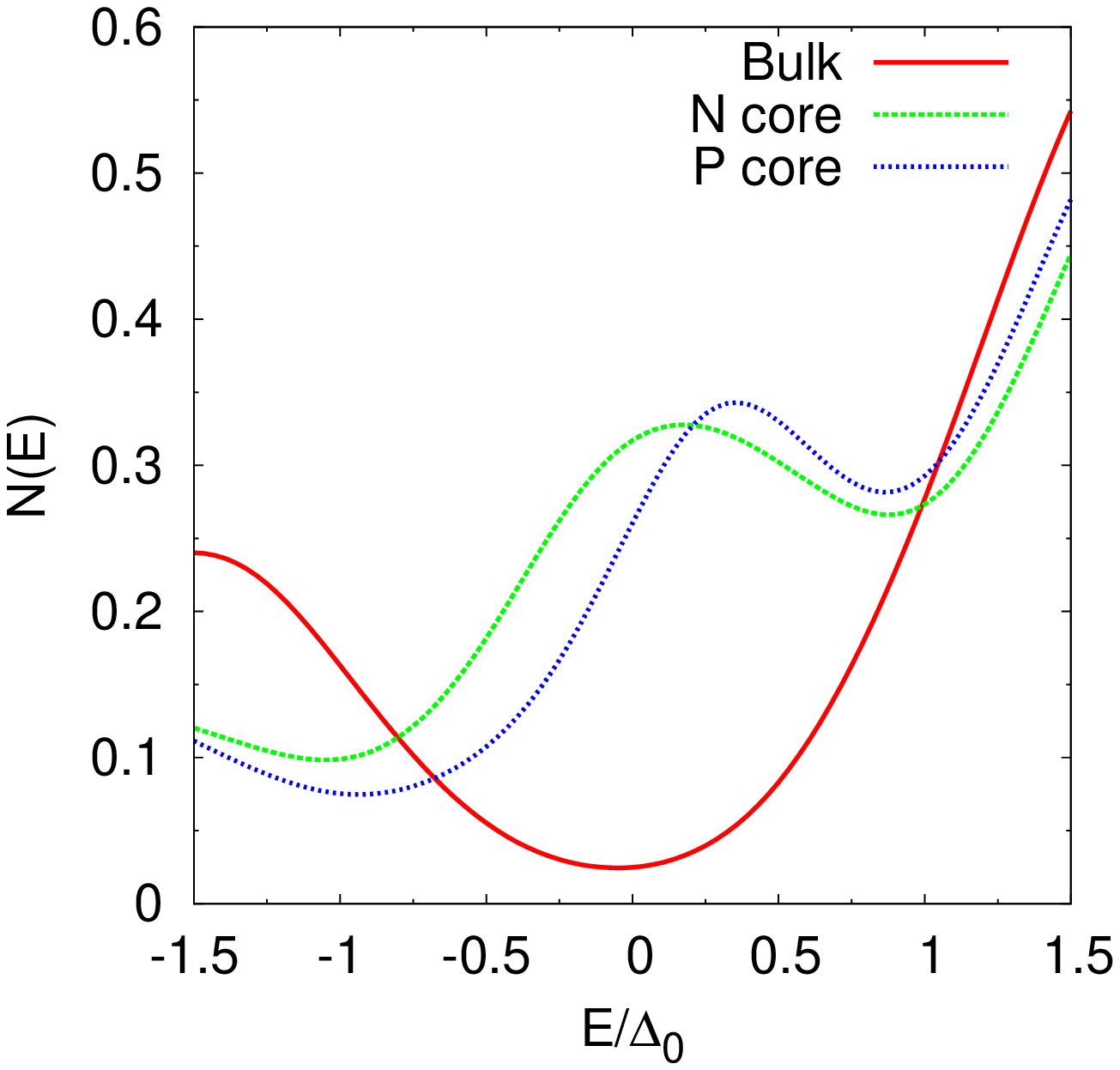}
        \put(5,80){\resizebox{25pt}{!}{(a)}}
      \end{overpic}} \\
    \resizebox{0.8\linewidth}{!}{
      \begin{overpic}{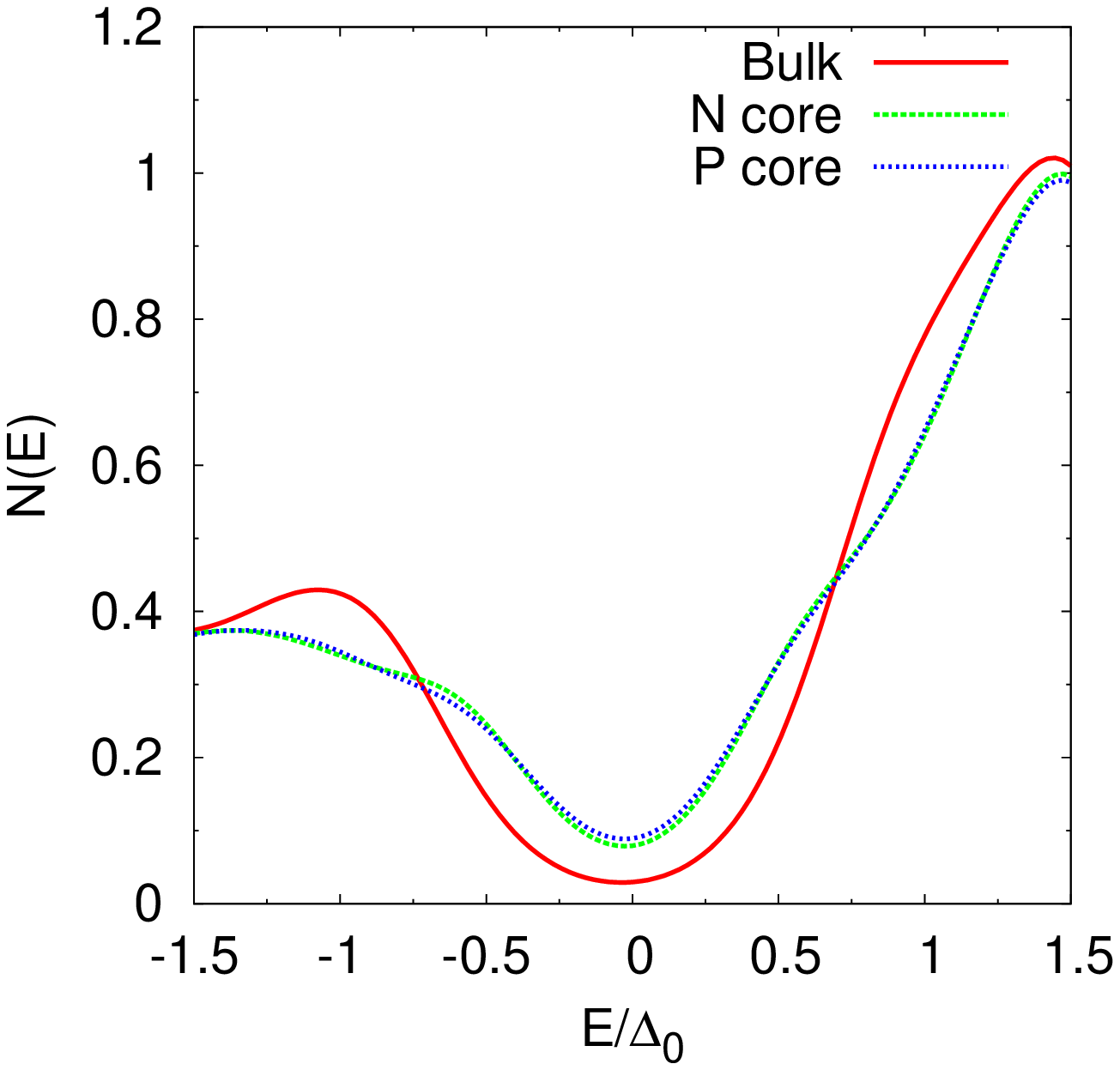}
        \put(5,80){\resizebox{25pt}{!}{(b)}}
      \end{overpic}}
  \end{tabular}
\caption{\label{dos}(Color online) LDOS $N(E,\bm{r})$ as a function of energy $E$ at various positions for (a) the single band model and (b) quasi-1D model. The bulk value is obtained in the absence of a magnetic field, whereas ``N core'' (``P core'') indicates the position at the negative (positive) vortex core. The temperature is about $0.3T_c$. }
\end{figure}

After having a general perspective on the vortex structure for the two different models, below we would like to study and analyse the LDOS in detail. The energy dependence of LDOS can be directly probed by measuring the tunneling conductance with an appropriate voltage bias in the STM experiment. In Fig.~\ref{dos}, we show a comparison of the energy-resolved LDOS at various positions for the two models. In the single band model, we can see the LDOS at the core of the negative vortex has a peak at $E\!=\!0$, while the peak for the positive vortex is slightly higher than zero. This feature, qualitatively consistent with previous calculations, is due to different winding structures of the negative and positive vortices \cite{Takigawa2001}.

However, the situation is totally different in the quasi-1D model. Similar study on organic superconductors in a magnetic field \cite{Takigawa2006} indicates that vortices are strongly modified due to the quasi-1D nature of superconductivity. Specifically, it has been demonstrated that vortices in a quasi-1D superconductor do not possess low energy mid-gap excitations. This extraordinary property leads to the missing of mid-gap resonance peak in the energy dependence of LDOS at the vortex site. To see whether this conclusion can be applied to our quasi-1D model for Sr$_2$RuO$_4$ or not, we also calculate the energy dependence of the LDOS at different positions in Fig.~\ref{dos} (b). It can be seen that the mid-gap resonance is also absent in the quasi-1D model we consider here. Nevertheless, we want to stress that this conclusion is valid only when the inter-orbital hopping amplitude $t'$ is small enough. On the contrary, if $t'$ is large enough, say $t'\!>\!0.4t$, we find that the mid-gap resonance will be present again, similar to the single band model.

\begin{figure}[htb!]\centering
  \begin{tabular}{c}
    \resizebox{0.8\linewidth}{!}{
      \begin{overpic}{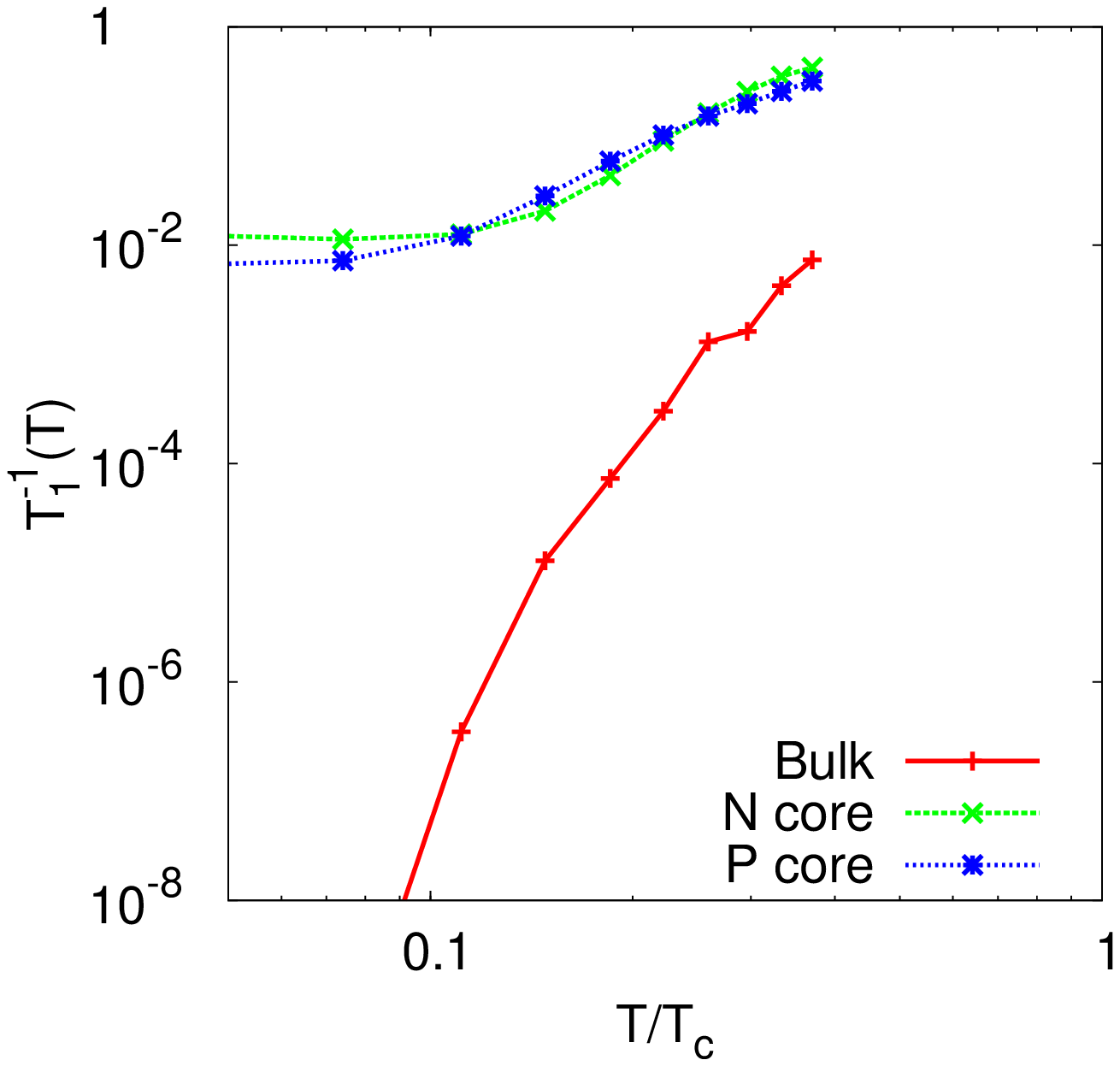}
        \put(5,80){\resizebox{25pt}{!}{(a)}}
      \end{overpic}} \\
    \resizebox{0.8\linewidth}{!}{
      \begin{overpic}{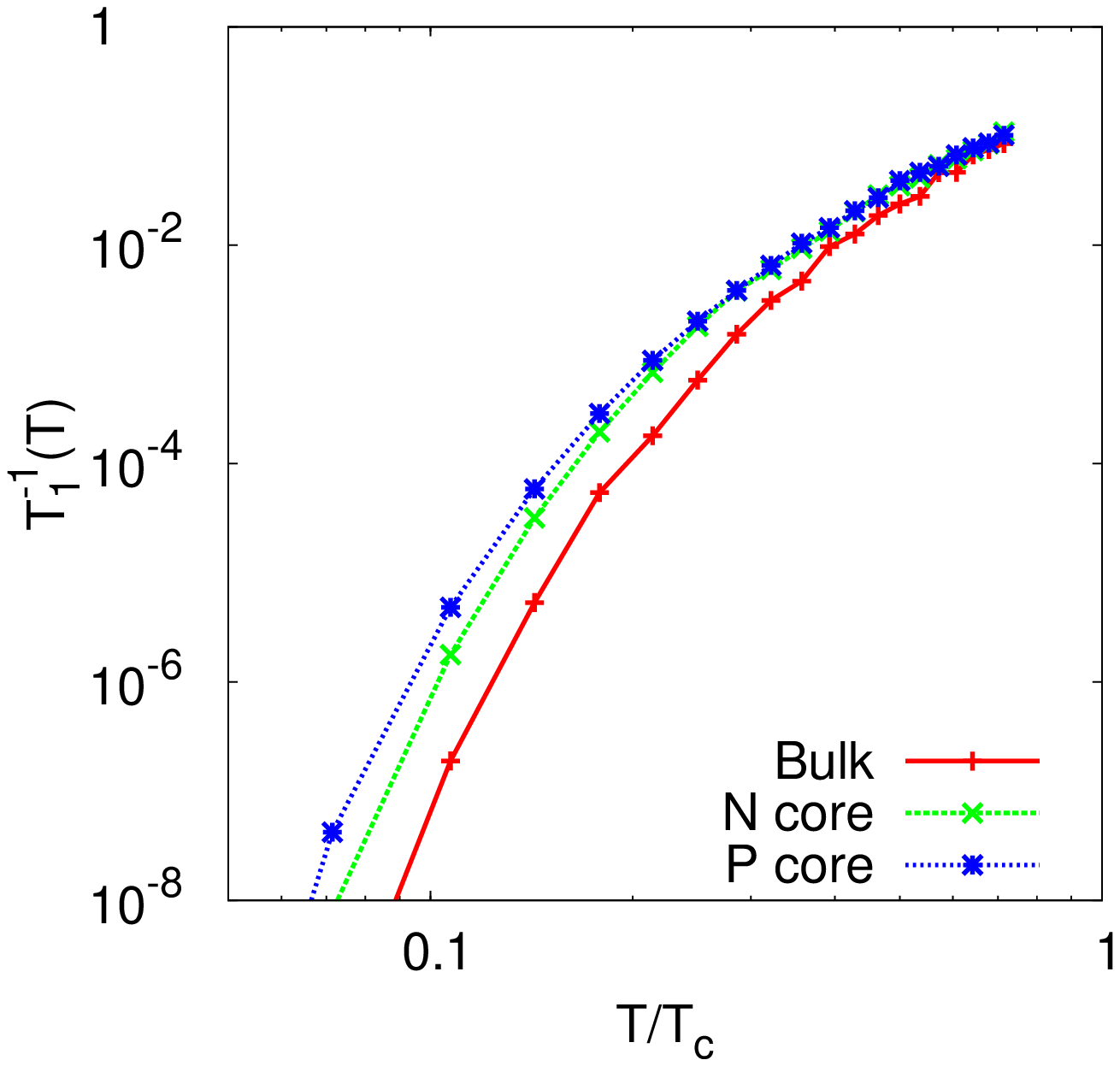}
        \put(5,80){\resizebox{25pt}{!}{(b)}}
      \end{overpic}}
  \end{tabular}
\caption{\label{nmr}(Color online) The nuclear spin-lattice relaxation rate $T_1^{-1}$ as a function of temperature $T$ at various positions for (a) a single band model and (b) a quasi-1D model. The bulk value is obtained in the absence of a magnetic field, whereas ``N core'' (``P core'') indicates the position at the negative (positive) vortex core.}
\end{figure}

Not only does the STM technique enable direct detection of the mid-gap excitations at the vortex core, but these excitations are also expected to be indicated in the NMR experiment. Now, we proceed to the discussion of the nuclear spin-lattice relaxation rate $T_1^{-1}$, which can be calculated via Eq.~\ref{eqn:nmr}. For the single band model, from Fig.~\ref{nmr} (a) we can see that a residual relaxation rate shows up at the vortex core as the temperature approaches zero, making $T^{-1}_1$ deviates from the bulk value in the absence of a magnetic field by several orders of magnitude especially at low temperatures, consistent with previous calculations~\cite{Takigawa2002}. This huge deviation stems from the fact that the mid gap states of the vortex core contribute substantially to the relaxation rate, compared to the zero field case. Next, we turn to the quasi-1D scenario in Fig.~\ref{nmr} (b). Note that the temperature dependence of the relaxation rate at different positions is almost identical in the logarithmic scale. The only subtle difference between the bulk and the vortex core is that $T_1^{-1}$ is slightly enhanced at low temperatures in the vortex core, but the previous residual behavior is totally absent. This result is consistent with the previous discussions on the energy dependence of LDOS, which also show similar behaviors between the bulk and the vortex core.

For Sr$_2$RuO$_4$, the NMR measurement has been performed experimentally in the absence of a magnetic field down to 0.1K \cite{Ishida2000}. And we suggest that further experiment in an external field should be performed as a crucial test of the two candidate models.

\section{Summary and Conclusion}\label{sec:conclusion}
We have studied the vortex state for a chiral $p$-wave superconductor based on a single 2D band and two quasi-1D band models, respectively. By comparing the two sets of results, we have found several distinctive characteristics in the vortex state derived from different models. Generally speaking, for the band structure of Sr$_2$RuO$_4$ at the Fermi surface, the $\gamma$ band is more or less isotropic in the $x\!-\!y$ plane, whereas the quasi-1D $\alpha$ and $\beta$ bands are highly anisotropic. Assuming  that superconductivity originates from either the $\gamma$ band or $\alpha$ and $\beta$ bands, the vortex state does inherit distinguishable features from the active band(s). Firstly, the shape of the LDOS at zero bias in the quasi-1D model shows anisotropy in accordance with the analysis on the angle-resolved Fermi velocity for the $\alpha$ and $\beta$ bands. Secondly, the missing of the mid-gap resonance at the vortex core for this model leads to corresponding consequences in the energy dependence of LDOS and the temperature dependence of nuclear spin-lattice relaxation rate $T_1^{-1}$. All these specialities show sharp distinction compared with the counterparts in the single-band scenario. Consequently, the STM and NMR measurements in the vortex state of Sr$_2$RuO$_4$ are expected to be unambiguous experiments to answer the question which band the superconductivity resides in. Before concluding, we would like to mention that it has been recently proposed that this disputatious issue related to the orbital origin of superconductivity can be settled by detecting the Leggett-like collective modes \cite{Chung2012}.

\begin{acknowledgments} 
We thank T. M. Rice for discussion. This work is partly in support by HK RGC GRF grant HKU10, and by NSFC grant number J20121499.
\end{acknowledgments}

\bibliography{mylib}
\bibliographystyle{apsrev4-1}
\end{document}